%% file: iclr2027_conference.tex
\documentclass{article} 
\usepackage{iclr2027_conference,times}

\input{math_commands.tex}

\usepackage{booktabs}
\usepackage{array}
\usepackage[table]{xcolor}
\usepackage{graphicx}
\usepackage{float}
\usepackage{placeins}
\usepackage{hyperref}
\usepackage{url}

\graphicspath{{figures/}}

\title{Learning Where to Split: Dynamic Band Splitting for Audio Source Separation}

\author{Bingzhe Zhang$^{1}$\footnotemark[1]\hphantom{$^{*}$}, Kai Li$^{1,2}$\thanks{Bingzhe Zhang and Kai Li contribute equally to the article.}\hphantom{$^{*}$}\footnotemark[2]\hphantom{$^{\dagger}$},
\& Xiaolin Hu$^{1,2,3}$\thanks{Corresponding authors.}\hphantom{$^{\dagger}$} \\
1. Department of Computer Science and Technology, Institute for AI, BNRist, \\
   Tsinghua University, Beijing 100084, China \\
2. IDG/McGovern Institute for Brain Research, Tsinghua University, Beijing 100084, China \\
3. Chinese Institute for Brain Research (CIBR), Beijing 100010, China \\
\texttt{\{zbz24,li-k24\}@mails.tsinghua.edu.cn} \\
\texttt{xlhu@tsinghua.edu.cn}
}

\definecolor{dbsgray}{gray}{0.93}
\newcommand{\dbsgain}[2]{#1\textsubscript{\textcolor{red}{+#2}}}

\iclrfinalcopy 
\begin{document}

\maketitle
\pagestyle{plain} 

\begin{abstract}
Audio source separation models must preserve fine spectral structure while keeping the cost of the separator tractable. Band-split architectures reduce this cost by manually partitioning the spectrum into predefined frequency bands and encoding each band as a token, but the same boundaries are used for every mixture. Because informative spectral structure varies with source content, fixed boundaries can merge discriminative components or spend tokens where little resolution is needed. We introduce Dynamic Band Splitting (DBS), an end-to-end framework that adapts boundary locations and band widths to each mixture under a fixed band-token budget. DBS predicts content-conditioned boundary scores within frequency blocks, selects a prescribed number of boundaries using hard Top-$K$, and applies content-dependent pooling to obtain compact band tokens. After separation, it smooths and expands these tokens to the original STFT grid, refines them along frequency, and decodes complex masks with a frequency-shared head. DBS improved separation over the corresponding static models on DnR v2, MUSDB18-HQ, and EchoSet, with DBS-BSRNN achieving relative gain of 29.0\% in average SI-SDR on DnR v2. These results and boundary analyses on DnR v2 show that content-adaptive frequency partitioning benefits multiple separator backbones under a fixed band-token budget. Code and audio demos are available at \url{https://cslikai.cn/Dynamic-Band-Splitting}.
\end{abstract}

\begin{figure*}[h]
    \centering
    \includegraphics[width=1.0\textwidth]{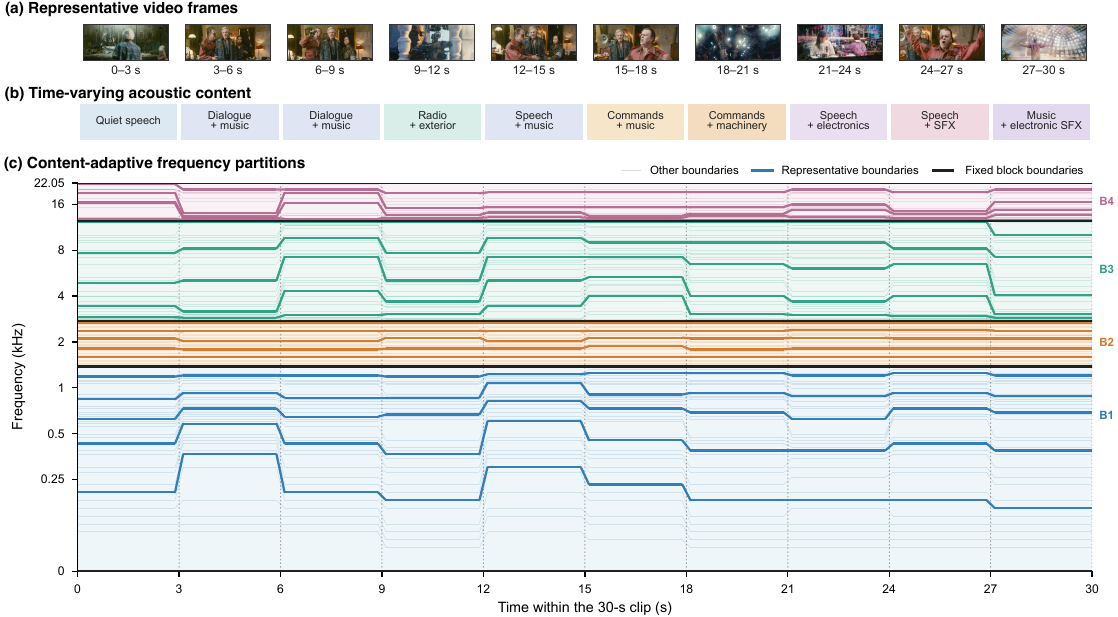}
    \caption{DBS partitions across a 30-second \emph{Tears of Steel} excerpt. Each column represents a non-overlapping three-second window, showing its midpoint frame, dominant acoustic content, and predicted partition. Colored lines highlight five learned boundaries per block; faint lines show the others, and black lines mark fixed block limits.}
    \label{fig:qualitative_teaser}
    \vspace{-15pt}
\end{figure*}


\section{Introduction}
\label{sec:introduction}

Audio source separation aims to recover individual sources, such as speech, musical instruments, and environmental sounds, from a recorded mixture. It supports music remixing and audio restoration \citep{rafii2018overview,stoter2019openunmix,li2026semantically}, hearing assistance \citep{li2025advances}, and meeting transcription and automatic subtitle generation \citep{raj2021meeting}. These applications require representations that capture source-specific acoustic structure across heterogeneous mixtures.

Source separation methods encode a mixture into acoustic features, model dependencies within these features, and reconstruct the constituent sources. Representative methods include convolutional models such as Conv-TasNet \citep{luo2019convtasnet}, encoder--decoder networks such as Demucs \citep{defossez2019demucs}, and attention-guided models such as TDANet \citep{li2023tdanet}. Alongside advances in separator design, band splitting compresses the spectral features supplied to the separator. BSRNN \citep{luo2023bsrnn} first represents the mixture as a complex spectrogram using the short-time Fourier transform (STFT), then uses a manually specified non-uniform layout to group its frequency bins into bands and map each band to a fixed-dimensional representation before modeling temporal and inter-band dependencies. Subsequent architectures, including BS-RoFormer \citep{lu2024bsroformer}, Mel-Band RoFormer \citep{wang2023melband}, and TIGER \citep{xu2025tiger}, also construct compact band representations using predefined frequency layouts. This reduces the frequency-axis sequence from individual STFT bins to a smaller set of bands, lowering the cost of subsequent separation.

However, these band-split methods apply the same frequency layout to every mixture \citep{luo2023bsrnn,lu2024bsroformer,xu2025tiger}. Their allocation of frequency resolution is therefore determined in advance, even though the locations and density of informative spectral components vary with the sources present. A wide band may compress several discriminative components into a single representation, while finely divided bands elsewhere may cover less informative content. Increasing the number of bands provides finer resolution but also increases the computational cost of the separator. The resulting challenge is to adapt the allocation of frequency resolution to each mixture while keeping the number of band representations fixed.

We propose Dynamic Band Splitting (DBS), an end-to-end framework that learns mixture-dependent frequency partitions under a fixed band-token budget. DBS divides the spectrogram's frequency axis into fixed frequency blocks and predicts content-dependent boundary scores within each block. Hard Top-$K$ selection forms a prescribed number of contiguous bands per block, allowing their locations and widths to change across mixtures. After separation, the band features are smoothed in token order and expanded to the original STFT grid for full-resolution reconstruction. This design decouples the frequency resolution used by the separator from that of its output and supports integration with different backbones, including BSRNN \citep{luo2023bsrnn} and TIGER \citep{xu2025tiger}. Figure~\ref{fig:qualitative_teaser} illustrates this behavior on a continuous 30-second movie excerpt: the block scaffold and 128-band budget remained fixed, while the learned boundaries changed with the acoustic content.

We evaluated DBS on the DnR v2 benchmark for music, speech, and sound-effect separation \citep{petermann2022cocktail}, MUSDB18-HQ for music separation \citep{rafii2019musdb18hq}, and EchoSet for speech separation \citep{xu2025tiger}. Relative to the BSRNN result, DBS-BSRNN improved average SI-SDR on DnR v2 by 29.0\%. DBS-BSRNN improved four-stem SDR over the static BSRNN model on MUSDB18-HQ by 6.8\% and improved the reported BSRNN SDRi on EchoSet by 3.6\%. The gains extended to other band-split backbones: BS-RoFormer \citep{lu2024bsroformer} and MelRoFormer \citep{wang2023melband} outperformed their static counterparts in four-stem SDR on MUSDB18-HQ, while DBS-TIGER exceeded the TIGER SDRi on EchoSet. Boundary analyses and fixed-boundary controls on DnR v2 showed that DBS adapted its partitions to the input and improved separation under a fixed band-token budget.


\section{Related Work}
\label{sec:related_work}

\textbf{Band-Split-Based Audio Source Separation.}
Time--frequency-domain source separation methods such as Open-Unmix~\citep{stoter2019openunmix} and TF-GridNet~\citep{wang2023tf} make spectral structure explicit, but processing high-resolution spectrograms over the full frequency axis can be computationally expensive. Band-split-based source separation methods address this cost by dividing the spectrum into subbands and assigning specialized modules to model within-band structure and cross-band interactions. Band-Split RNN (BSRNN)~\citep{luo2023bsrnn}, BS-RoFormer~\citep{lu2024bsroformer}, Mel-Band RoFormer~\citep{wang2023melband}, and TIGER~\citep{xu2025tiger} instantiate this design with different subband representations and cross-band modeling modules. SCNet~\citep{tong2024scnet} integrates subband-specific compression into a hierarchical encoder--decoder, coupling frequency downsampling with corresponding upsampling stages and skip connections. Across these designs, the frequency layout is treated as a model-level structural prior: boundaries and band widths are specified by design and shared across inputs.

\textbf{Adaptive Spectral Representation and Frequency Allocation.}
Learnable filterbanks and input-conditioned modules relax the fixed spectral front end in different ways. Multi-Channel Masking with Learnable Filterbank~\citep{dai2023learnablefilterbank} optimizes the spectral front end during training, whereas Spectral Feature Compression (SFC)~\citep{saijo2026sfc} replaces the fixed band-split encoder with a shared sequence-modeling module based on cross-attention or Mamba. Sparsely gated mixture-of-experts source separation models~\citep{wang2022sparsemoeseparation} have been used to balance separation quality, speech distortion, and computational cost, while TF-MoE~\citep{hu2026tfmoe} routes experts along the time and frequency axes. These methods adapt the spectral representation or computational path within a particular separation model. DBS instead learns input-dependent boundaries between contiguous frequency bands while keeping the number of bands fixed. Its boundary predictor is optimized jointly with the separator, and the original STFT grid is retained for reconstruction.


\section{Method}
\label{sec:method}

\begin{figure}[h]
    \centering
    \includegraphics[pagebox=cropbox,width=\linewidth]{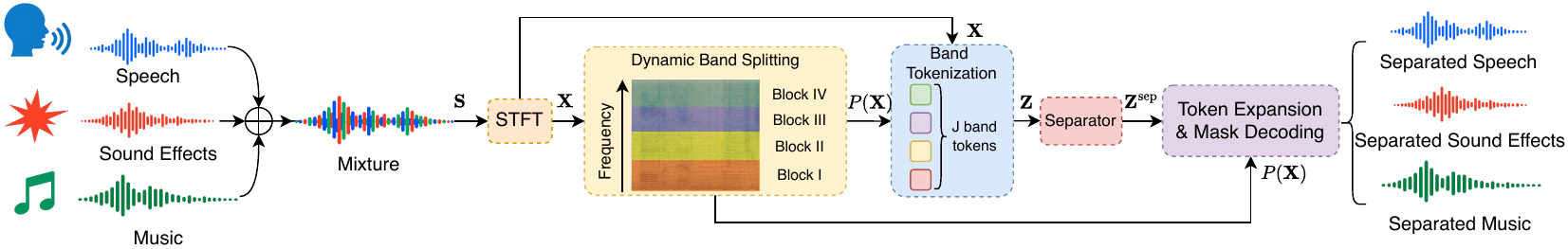}
    \caption{
        Overview of DBS, using cinematic audio separation on the DnR v2 dataset as an example.
    }
    \label{fig:dbs_overview}
\end{figure}

Let $\mathbf{x}\in\mathbb{R}^{L}$ denote a monaural mixture and $\mathbf{X}=\operatorname{STFT}(\mathbf{x})\in\mathbb{C}^{F\times T}$ its complex spectrogram, with $F$ frequency bins and $T$ time frames. The goal is to estimate $K$ source waveforms $\{\hat{\mathbf{s}}_k\in\mathbb{R}^{L}\}_{k=1}^{K}$. DBS operates on the frequency bins of $\mathbf{X}$: it predicts a mixture-dependent partition $P(\mathbf{X})$, pools its $J$ bands into a fixed-length token sequence, and processes the tokens with a separation backbone. The same partition then maps separated tokens back to the STFT grid for complex-mask estimation (Figure~\ref{fig:dbs_overview}). Sections~\ref{sec:dynamic_partition}--\ref{sec:full_resolution_reconstruction} describe these stages, followed by end-to-end optimization in Section~\ref{sec:optimization}.

\subsection{Dynamic Band Splitting}
\label{sec:dynamic_partition}

\begin{figure}[h]
    \centering
    \includegraphics[pagebox=cropbox,width=\linewidth]{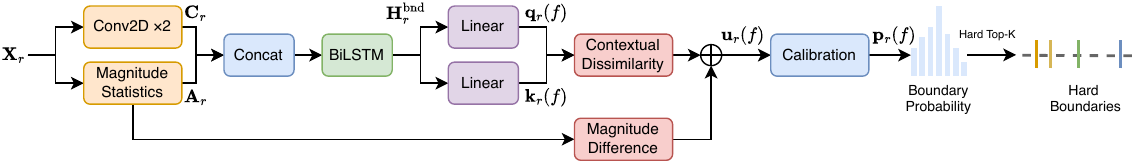}
    \caption{Detailed dynamic band splitting within block $r$.}
    \label{fig:dbs_boundary_predictor}
\end{figure}

Fixed band layouts assign the same frequency resolution to every mixture. DBS instead moves band boundaries according to the input while keeping the separator's token count constant. We divide the frequency axis of $\mathbf{X}$ into $R$ fixed, contiguous, non-overlapping frequency blocks $\{\mathcal{F}_r\}_{r=1}^{R}$, with $F_r=|\mathcal{F}_r|$ bins in block $r$, and assign that block a budget of $J_r$ bands. The total budget is $J=\sum_{r=1}^{R}J_r$, which is identical for every mixture. The blocks ensure that each part of the spectrum receives a prescribed share of the budget, while boundaries within each block remain adaptive. We used four blocks (B1--B4), a choice examined in Section~\ref{sec:design_ablation}. Figure~\ref{fig:dbs_boundary_predictor} details how the internal boundaries are selected within each block.

For $\mathbf{X}_r=\mathbf{X}[\mathcal{F}_r,:]\in\mathbb{C}^{F_r\times T}$, the predictor processes its real and imaginary components with temporal convolutions and extracts three magnitude statistics per frequency bin: temporal mean, peak-to-mean ratio, and standard-deviation-to-mean ratio. A BiLSTM summarizes the fused features into a contextual vector $\mathbf{h}_r(f)\in\mathbb{R}^{d}$ for each bin. For each internal candidate $f>f_r^{\min}$, representing the transition from bin $f-1$ to bin $f$, the score has two components: $u_r(f)=u_r^{\mathrm{feat}}(f)+0.15u_r^{\mathrm{mag}}(f)$. The feature term compares learned representations of the adjacent bins, while the magnitude term measures the change in their acoustic statistics after a bounded transform. A learned scale $\tau_r$ and position bias $\beta_r(f)$ then give the boundary probability
\begin{equation}
    p_r(f)=\sigma\!\left(\tau_r\bigl(u_r(f)-0.5\bigr)+\beta_r(f)\right).
\end{equation}
Appendix~\ref{app:boundary_predictor} specifies the feature branches, score terms, and calibration parameters.

The probability field $p_r(f)$ provides candidate boundaries, but the realized partition must contain exactly $J_r$ bands. We fix the first bin $f_r^{\min}=\min\mathcal{F}_r$ as a boundary and use Hard Top-$K$ to select the remaining $J_r-1$ positions:
\begin{equation}
    \mathcal{B}_r
    =
    \{f_r^{\min}\}
    \cup
    \operatorname{TopK}
    \left(
        \{p_r(f)\}_{f\in\mathcal{F}_r\setminus\{f_r^{\min}\}},
        J_r-1
    \right),
\end{equation}
Here, $\operatorname{TopK}$ returns the positions of the largest candidate probabilities. Selecting $J_r-1$ of the $F_r-1$ internal transitions yields $J_r$ contiguous bands because the first bin starts the first band. Let $b_r(f)\in\{0,1\}$ indicate whether $f\in\mathcal{B}_r$. Cumulative counting assigns each bin to a band:
\begin{equation}
    s_r(f)
    =
    \sum_{i=f_r^{\min}}^{f} b_r(i),
    \qquad s_r(f)\in\{1,\ldots,J_r\}.
\end{equation}
The assignments $\{s_r(f)\}$ define the mixture-dependent partition $P(\mathbf{X})$ passed to tokenization. Their locations can change without changing the $J$-band budget.


\subsection{Band Tokenization and Separation}
\label{sec:dynamic_tokenization}
\label{sec:shared_modeling}

\textbf{Band Tokenization.}
The bands selected in Section~\ref{sec:dynamic_partition} have variable widths, whereas the separator expects a fixed feature dimension per token. The hard assignment $s_r(f)$ collects the bins of band $j$ in block $r$:
\begin{equation}
    \mathcal{S}_{r,j}
    =
    \{f\in\mathcal{F}_r:s_r(f)=j\},
    \qquad j=1,\ldots,J_r.
\end{equation}
We stack the real and imaginary components of $\mathbf{X}_r$ and encode each bin as $\mathbf{U}_r(f,t)\in\mathbb{R}^{D}$. A learned score of its time-averaged feature determines a pooling weight $\alpha_{r,j}(f)$, normalized over bins in $\mathcal{S}_{r,j}$. Weighted pooling produces one token per band:
\begin{equation}
    \mathbf{Z}_{r,j}(t)
    =
    \sum_{f\in\mathcal{S}_{r,j}}
    \alpha_{r,j}(f)
    \mathbf{U}_r(f,t)
    \in\mathbb{R}^{D}.
\end{equation}
The weights are shared across time frames, so every band contributes one $D$-dimensional token at each frame regardless of its width. Concatenating bands in frequency order gives $\mathbf{Z}\in\mathbb{R}^{J\times T\times D}$. Thus, DBS changes which frequency bins are pooled together while preserving the number $J$ and feature dimension $D$ of separator inputs. Appendix~\ref{app:tokenization_details} specifies the bin encoder and pooling scores.

\textbf{Separator.}
The fixed-size token sequence $\mathbf{Z}$ provides a common input to band-split separation backbones. We instantiated DBS with BSRNN, TIGER, BS-RoFormer, and MelRoFormer (Section~\ref{sec:main_results}). Each backbone modeled temporal and inter-band dependencies and produced separated features $\mathbf{Z}^{\mathrm{sep}}\in\mathbb{R}^{J\times T\times D_{\mathrm{sep}}}$. Tokens from all blocks enter the same sequence, allowing cross-block interactions. Their band order is retained so that the decoder can use $P(\mathbf{X})$ to restore frequency resolution.


\subsection{Token Expansion \& Mask Decoding}
\label{sec:full_resolution_reconstruction}

\begin{figure}[h]
    \centering
    \includegraphics[pagebox=cropbox,width=\linewidth]{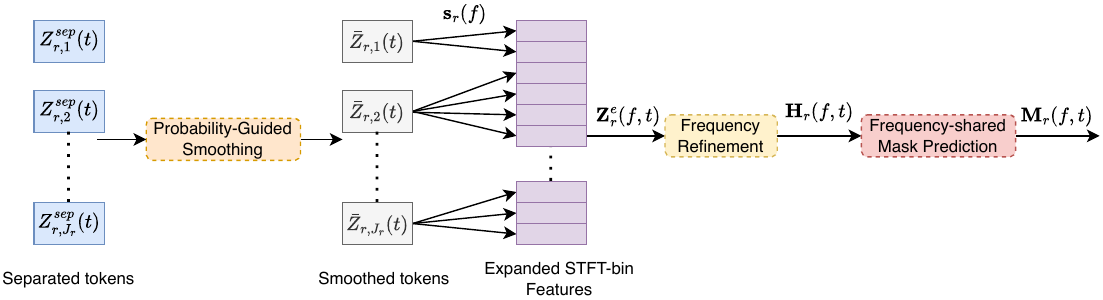}
    \caption{
        Full-resolution reconstruction within block $r$.
    }
    \label{fig:dbs_reconstruction}
\end{figure}

The separated representation $\mathbf{Z}^{\mathrm{sep}}$ contains one feature vector per band and time frame, whereas source reconstruction requires a mask at every STFT bin. We therefore smooth the separated tokens, expand them with the same hard partition used by the encoder, and refine the expanded features before predicting complex masks. Figure~\ref{fig:dbs_reconstruction} summarizes this path.

\textbf{Probability-Guided Smoothing.}
Expanding unsmoothed tokens would create piecewise-constant features with jumps at band boundaries. We average the boundary probabilities within each selected band $\mathcal{S}_{r,j}$ to obtain $\bar p_{r,j}$ and use it to interpolate the separated token $\mathbf{Z}^{\mathrm{sep}}_{r,j}(t)$ with the preceding smoothed token $\bar{\mathbf{Z}}_{r,j-1}(t)$. Higher $\bar p_{r,j}$ retains more of the current token, whereas lower values pass more information from the preceding band. This can soften feature transitions while preserving the selected boundaries and token count. The exact interpolation is given in Appendix~\ref{app:reconstruction_details}.

\textbf{Token Expansion.}
The smoothed tokens still have band-level resolution. To recover a feature at each frequency bin, we use the hard assignment $s_r(f)$ from Section~\ref{sec:dynamic_partition} to copy the corresponding token to that bin:
\begin{equation}
    \mathbf{Z}_r^{e}(f,t)
    =
    \bar{\mathbf{Z}}_{r,s_r(f)}(t)
    \in\mathbb{R}^{D_{\mathrm{sep}}}.
    \label{eq:hard_token_to_bin}
\end{equation}
This yields a feature map $\mathbf{Z}_r^{e}\in\mathbb{R}^{F_r\times T\times D_{\mathrm{sep}}}$ on the original STFT grid.

\textbf{Frequency Refinement.}
The expanded map $\mathbf{Z}_r^{e}$ assigns the same feature to all bins in a band. A residual frequency-direction block produces $\mathbf{H}_r=\mathcal{R}_r^{\mathrm{freq}}(\mathbf{Z}_r^{e})\in\mathbb{R}^{F_r\times T\times D_{\mathrm{sep}}}$ by mixing neighboring bin features, including across band boundaries. Appendix~\ref{app:reconstruction_details} specifies its depthwise and pointwise operations.

\textbf{Frequency-shared Mask Prediction.}
Within block $r$, the frequency-shared decoder maps each refined feature $\mathbf{H}_r(f,t)$ to mask parameters $\boldsymbol{\Theta}_r(f,t)=\mathcal{D}_r(\mathbf{H}_r(f,t))\in\mathbb{R}^{K\times2\times2}$. These parameters produce gated real and imaginary mask components for each source. After a source-axis consistency projection when $K>1$, we concatenate the complex masks $\mathbf{M}_{r,k}\in\mathbb{C}^{F_r\times T}$ from all blocks along frequency to obtain $\mathbf{M}_k\in\mathbb{C}^{F\times T}$ (Appendix~\ref{app:reconstruction_details}). The estimated source spectrogram is
\begin{equation}
    \hat{\mathbf{S}}_k
    =
    \mathbf{X}\odot\mathbf{M}_k
    \in\mathbb{C}^{F\times T}.
\end{equation}
The inverse STFT transforms $\hat{\mathbf{S}}_k$ into the waveform $\hat{\mathbf{s}}_k$.


\subsection{End-to-End Training}
\label{sec:optimization}

The hard boundary decisions in Section~\ref{sec:dynamic_partition} fix the forward token sequence but block ordinary gradients from the separation loss to the boundary predictor. Following the straight-through estimator~\citep{bengio2013estimating}, we use the selected hard partition in the forward pass and differentiate through a local soft assignment in the backward pass. For bin $f$, let $\mathbf{Z}_r^{e}(f)$ collect its hard-expanded features over time and let $\mathbf{Z}_r^{\mathrm{soft}}(f)$ interpolate the features obtained when a boundary is present or absent, weighted by $p_r(f)$. The training representation is
\begin{equation}
    \mathbf{Z}_r^{\mathrm{STE}}(f)
    =\mathbf{Z}_r^{e}(f)+\mathbf{Z}_r^{\mathrm{soft}}(f)
    -\operatorname{sg}\!\left(\mathbf{Z}_r^{\mathrm{soft}}(f)\right)
    \in\mathbb{R}^{T\times D_{\mathrm{sep}}},
    \label{eq:training_ste}
\end{equation}
Here, $\operatorname{sg}$ stops gradients, so the forward value equals $\mathbf{Z}_r^{e}(f)$ while the backward pass follows the local surrogate. The surrogate gradient to $p_r(f)$ depends on the difference between the boundary-present and boundary-absent features. A value-preserving confidence path supplies another gradient to the boundary probabilities without changing the forward features. Appendix~\ref{app:ste} gives both constructions.

To control the soft boundary density, let $\rho_r>0$ denote the target compression ratio in block $r$ and define
\begin{equation}
    \mathcal{L}_{\mathrm{ratio}}
    =\frac{1}{R}\sum_{r=1}^{R}
    \left(\frac{1}{F_r}\sum_{f\in\mathcal{F}_r}p_r(f)
    -\frac{1}{\rho_r}\right)^2,
    \label{eq:ratio_loss}
\end{equation}
where $1/\rho_r$ is the target mean boundary probability. The joint objective is
\begin{equation}
    \mathcal{L}
    =
    \mathcal{L}_{\mathrm{sep}}
    +
    \lambda_{\mathrm{ratio}}
    \mathcal{L}_{\mathrm{ratio}},
\end{equation}
where $\lambda_{\mathrm{ratio}}\geq0$ is the regularization weight and $\mathcal{L}_{\mathrm{sep}}$ is the task-specific source-separation loss. Through the straight-through path, $\mathcal{L}_{\mathrm{sep}}$ updates the boundary predictor alongside the separator and decoder; $\mathcal{L}_{\mathrm{ratio}}$ regularizes the soft boundary probabilities, while Hard Top-$K$ fixes the realized band count at $J$. Appendix~\ref{app:separation_objectives} gives the loss definitions used in our experiments.

%
%
%

\section{Experiments}
\label{sec:experiments}


\subsection{Experimental Setup}
\label{sec:experimental_setup}

\textbf{Datasets.} We evaluated DBS on DnR v2~\citep{petermann2022cocktail}, MUSDB18-HQ~\citep{rafii2019musdb18hq}, and EchoSet~\citep{xu2025tiger}, covering cinematic, music, and speech source separation. DnR v2 contains music, speech, and sound effects; MUSDB18-HQ provides vocals, drums, bass, and other stems; and EchoSet contains two-speaker mixtures with noise and reverberation. At the sample rates used here, DnR v2 and MUSDB18-HQ had a 0--22.05~kHz analyzed frequency range, whereas EchoSet had a 0--8~kHz range. This contrast let us examine adaptive band allocation across different spectral bandwidths. Dataset construction and split sizes are detailed in Appendix~\ref{app:datasets}.

\textbf{Compared systems.} On DnR v2, we trained fixed-band and DBS variants of BSRNN and TIGER from scratch, each with 128 bands. Appendix~\ref{app:fixed_128_construction} details how each backbone's original partition was expanded to 128 bands. On MUSDB18-HQ, we trained fixed-band and DBS variants of BSRNN, BS-RoFormer, and MelRoFormer as single models that jointly estimated all four stems. This differed from the target-specific models used by the original BSRNN and BS-RoFormer systems~\citep{luo2023bsrnn,lu2024bsroformer}. All MUSDB models used only the MUSDB18-HQ training split, without external audio, and were implemented in the Music-Source-Separation-Training framework.\footnote{\url{https://github.com/ZFTurbo/Music-Source-Separation-Training}} On EchoSet, we compared DBS-BSRNN and DBS-TIGER with the systems reported by TIGER~\citep{xu2025tiger}. Appendix~\ref{app:backbone_configs} details the backbone substitutions.

\textbf{Training and evaluation.} DBS used four frequency blocks and a fixed band-token budget. DnR v2 and MUSDB18-HQ used 1025 STFT bins and 128 band tokens, whereas EchoSet used 321 bins and 67 band tokens. Each inference window determined one partition shared by band encoding and full-resolution reconstruction. We reported source-wise SI-SDR and its unweighted average on DnR v2, four-stem SDR on MUSDB18-HQ, and SDRi and SI-SDRi on EchoSet. Detailed optimizer settings, band construction, STFT parameters, losses, and training schedules are given in Appendix~\ref{app:implementation}; metric definitions and evaluation protocols are provided in Appendix~\ref{app:evaluation_protocol}.


\subsection{Separation Performance Across Datasets and Backbones}
\label{sec:main_results}

\textbf{DnR v2.} To retain fine spectral detail over DnR v2's broad 0--22.05~kHz frequency range, we use 128 bands for both the static and DBS variants. Table~\ref{tab:dnr_main} shows that DBS improves both BSRNN and TIGER over their static 128-band counterparts. Average SI-SDR increases from 7.69 to 9.07~dB for BSRNN and from 8.74 to 10.13~dB for TIGER, yielding gains of 1.38 and 1.39~dB, respectively. Both models improve across all three sources.

\begin{table*}[t]
    \centering
    \caption{Source separation performance on the DnR v2 test set, measured by SI-SDR (dB). Models marked with $\dagger$ are trained from scratch with 128 bands. Red subscripts indicate improvements over the corresponding static baselines. MACs are reported per second of input audio.}
    \label{tab:dnr_main}
    \footnotesize
    \setlength{\tabcolsep}{2pt}
    \resizebox{\textwidth}{!}{%
        \begin{tabular}{lrrrrrr}
            \toprule
            Method
            & Music $\uparrow$
            & Speech $\uparrow$
            & SFX $\uparrow$
            & Avg. $\uparrow$
            & Params (M) $\downarrow$
            & MACs (G/s) $\downarrow$ \\
            \midrule
            Conv-TasNet~\citep{luo2019convtasnet}
            & 0.30 & 8.50 & 2.00 & 3.60 & 5.3 & 19.8 \\
            MRX~\citep{petermann2022cocktail}
            & 4.20 & 12.30 & 5.70 & 7.40 & 30.5 & 10.6 \\
            BSRNN~\citep{luo2023bsrnn}
            & 5.50 & 13.80 & 1.80 & 7.03 & 52.8 & 18.2 \\
            TIGER~\citep{xu2025tiger}
            & 5.14 & 13.51 & 3.66
            & 7.44 & 1.4 & 10.8 \\
            \midrule
            BSRNN (128)$^{\dagger}$~\citep{luo2023bsrnn}
            & 4.83 & 14.87 & 3.38 & 7.69 & 35.5 & 76.6 \\
            \rowcolor{dbsgray}
            DBS-BSRNN (128)$^{\dagger}$
            & \dbsgain{6.48}{1.65}
            & \dbsgain{15.01}{0.14}
            & \dbsgain{5.73}{2.35}
            & \dbsgain{9.07}{1.38}
            & 4.7 & 124.0 \\
            \midrule
            TIGER (128)$^{\dagger}$
            & 6.18 & 14.91 & 5.13 & 8.74 & 1.4 & 20.4 \\
            \rowcolor{dbsgray}
            DBS-TIGER (128)$^{\dagger}$
            & \dbsgain{7.78}{1.60} & \dbsgain{15.90}{0.99}
            & \dbsgain{6.70}{1.57} & \dbsgain{10.13}{1.39} & 1.0 & 31.6 \\
            \bottomrule
        \end{tabular}%
    }
    \vspace{-10pt}
\end{table*}

\textbf{MUSDB18-HQ.} Table~\ref{tab:musdb_main} shows that DBS improves average four-stem SDR across all three backbones in the single-model setting. For BSRNN, average SDR increases from 5.09 to 5.43~dB. BS-RoFormer achieves the largest gain of 1.84~dB, improving from 4.50 to 6.34~dB. MelRoFormer improves from 7.25 to 8.07~dB, a gain of 0.82~dB. These results demonstrate the applicability of DBS across BSRNN, BS-RoFormer, and MelRoFormer.

\begin{table*}[t]
    \centering
    \caption{Music source separation performance on the MUSDB18-HQ test set, measured by SDR (dB). All models ($\dagger$) are trained in this work without external data and jointly separate the four stems.}
    \label{tab:musdb_main}
    \footnotesize
    \setlength{\tabcolsep}{2.4pt}
    \resizebox{\textwidth}{!}{%
        \begin{tabular}{lrrrrrrr}
            \toprule
            Model & Voc. $\uparrow$ & Drums $\uparrow$ & Bass $\uparrow$
            & Other $\uparrow$ & Avg. $\uparrow$ & Params (M) $\downarrow$
            & MACs (G/s) $\downarrow$ \\
            \midrule
            BSRNN$^{\dagger}$~\citep{luo2023bsrnn} & 6.01 & 6.23 & 4.49 & 3.62 & 5.09 & 71.7 & 21.8 \\
            \rowcolor{dbsgray}
            DBS-BSRNN$^{\dagger}$ & \dbsgain{6.70}{0.69} & \dbsgain{6.44}{0.21}
            & \dbsgain{4.58}{0.09} & \dbsgain{4.01}{0.39} & \dbsgain{5.43}{0.35} & 21.7 & 20.8 \\
            \midrule
            BS-RoFormer$^{\dagger}$~\citep{lu2024bsroformer} & 5.14 & 5.76 & 3.95 & 3.17 & 4.50 & 378.5 & 292.7 \\
            \rowcolor{dbsgray}
            DBS-BS-RoFormer$^{\dagger}$ & \dbsgain{7.64}{2.50} & \dbsgain{7.72}{1.96}
            & \dbsgain{5.29}{1.34} & \dbsgain{4.72}{1.55} & \dbsgain{6.34}{1.84} & 33.9 & 290.5 \\
            \midrule
            MelRoFormer$^{\dagger}$~\citep{wang2023melband} & 9.20 & 8.29 & 6.05 & 5.48 & 7.25 & 98.7 & 171.7 \\
            \rowcolor{dbsgray}
            DBS-MelRoFormer$^{\dagger}$ & \dbsgain{10.02}{0.82} & \dbsgain{9.30}{1.01}
            & \dbsgain{6.96}{0.91} & \dbsgain{5.98}{0.50} & \dbsgain{8.07}{0.82} & 19.2 & 252.9 \\
            \bottomrule
        \end{tabular}%
    }
    \vspace{-10pt}
\end{table*}

\textbf{EchoSet.} DBS-BSRNN improves SDRi and SI-SDRi by 0.46 and 0.43~dB over the published BSRNN baseline, while DBS-TIGER achieves gains of 0.33 and 0.31~dB over the published TIGER baseline, respectively. These results extend the benefits of DBS to speech separation. Full results are provided in Appendix~\ref{app:echoset_results}.


\subsection{Isolating the Effect of Dynamic Boundaries}
\label{sec:ablations}

\begin{table}[h]
    \vspace{-10pt}
    \centering
    \caption{Comparison of frequency partitions and front ends on the DnR v2 test set using a BSRNN separator and 128 bands.}
    \label{tab:from_scratch_ablation}
    \small
    \setlength{\tabcolsep}{5pt}
    \begin{tabular}{lrrrr}
        \toprule
        Variant
        & Music $\uparrow$
        & Speech $\uparrow$
        & SFX $\uparrow$
        & Avg. $\uparrow$ \\
        \midrule
        BSRNN (128)~\citep{luo2023bsrnn}
        & 4.83 & 14.87 & 3.38 & 7.69 \\
        \midrule
        Uniform
        & 5.58 & 14.43 & 3.99 & 8.00 \\
        Mel bands \citep{wang2023melband}
        & 2.85 & 14.65 & 0.10 & 5.87 \\
        Learnable filterbank~\citep{dai2023learnablefilterbank}
        & -11.47 & 8.62 & -32.30 & -11.72 \\
        \textbf{DBS-BSRNN}
        & \textbf{6.48} & \textbf{15.01} & \textbf{5.73}
        & \textbf{9.07} \\
        \bottomrule
    \end{tabular}
\end{table}

We used DBS-BSRNN on DnR v2 to assess input-conditioned boundary placement under the four-block, 128-band setting. The comparison included a uniform control, Mel bands, and a learnable filterbank, each with 128 bands.

Under the same architecture and training schedule, DBS-BSRNN improved average SI-SDR by 1.07~dB over the separately trained uniform control (Table~\ref{tab:from_scratch_ablation}). Gains extended to all three sources. Mel bands remained close to DBS on speech but lagged on music and sound effects. An unconstrained learnable filterbank baseline performed poorly in our setting; analysis in Appendix~\ref{app:learnable_frontend_diagnostics} links the degradation to the analysis--synthesis transform and its integration with the ordered band-split architecture. The gain over the matched fixed-boundary control supports adapting frequency allocation to mixture content to accommodate these differing source requirements within the same 128-band budget.

Replacing the learned boundaries with the BSRNN layout at inference reduced average SI-SDR from 9.07 to 8.62~dB while keeping model weights unchanged (Appendix~\ref{app:boundary_substitution}).

\begin{figure}[h]
    \centering
    \includegraphics[width=1.0\linewidth]{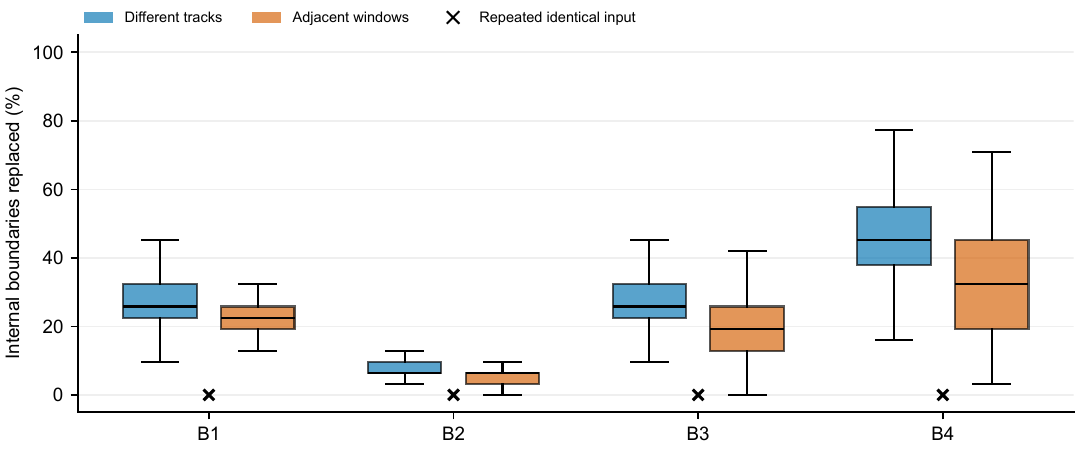}
    \caption{Boundary adaptation on DnR v2 under a fixed 128-band budget. Replacement is the percentage of internal boundaries not shared by two partitions of the same block.
    Crosses mark repeated inference on identical inputs. Center lines mark medians, boxes span interquartile ranges, and whiskers extend to the most extreme observations within 1.5 interquartile ranges of the quartiles.}
    \label{fig:boundary_dynamics_budget}
\end{figure}

\textbf{Boundary adaptation.} We next examined how the learned boundaries varied across mixtures and over time, using 64 DnR v2 test mixtures under the same 128-band budget. We compared central three-second windows across mixtures, adjacent windows within each recording, and repeated inference on identical inputs. All 64 central windows yielded distinct partitions, and boundaries also changed between adjacent windows (Figure~\ref{fig:boundary_dynamics_budget}). Repeated inference produced identical boundaries. These results showed input-dependent frequency allocation across mixtures and within recordings without changing the token count. Appendix~\ref{app:boundary_analysis} provides the sampling protocol, per-block statistics, and further visualizations.

\subsection{Design Ablations}
\label{sec:design_ablation}

We examine the roles of budget allocation across blocks, boundary prediction, and frequency reconstruction using DBS-BSRNN on DnR v2 with a fixed budget of 128 bands.

\begin{figure}[h]
    \centering
    \includegraphics[width=1.0\columnwidth]{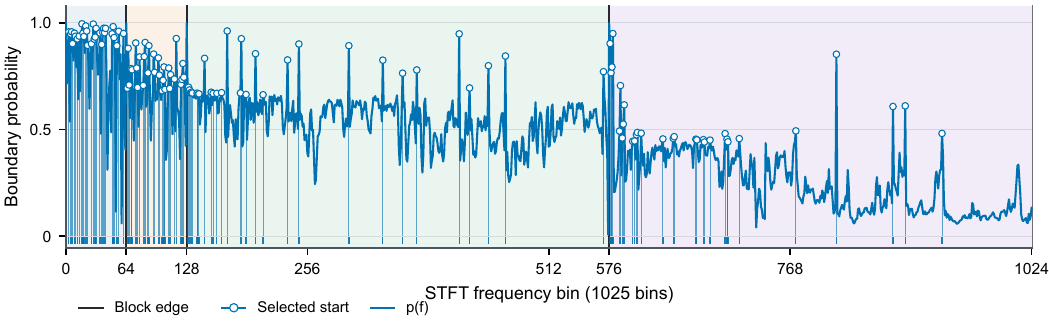}
    \caption{Boundary probabilities $p(f)$ and selected boundaries for the 1Block variant on a DnR v2 test window. Most selected boundaries fall at low frequencies, leaving the upper frequency range sparsely partitioned.}
    \label{fig:1block_boundary_result}
\end{figure}

\textbf{Block structure.} To test whether per-block budgets improve frequency allocation, we compare layouts with four, two, and one block while keeping the total band count fixed (Table~\ref{tab:block_ablation}). The two-block control merges B1 with B2 and B3 with B4. Average SI-SDR decreases from 9.07~dB with four blocks to 8.32~dB with two and 6.30~dB with one. In the one-block example in Figure~\ref{fig:1block_boundary_result}, selected boundaries concentrate at low frequencies, leaving the upper spectrum sparsely partitioned. These results support using per-block budgets to distribute boundaries across the spectrum and improve separation at the same token count.

\begin{table}[h]
    \centering
    \caption{Effect of the number of frequency blocks on DBS-BSRNN performance on the DnR v2 test set.}
    \label{tab:block_ablation}
    \small
    \setlength{\tabcolsep}{4pt}
    \begin{tabular}{lrrrr}
        \toprule
        Variant & Music $\uparrow$ & Speech $\uparrow$ & SFX $\uparrow$ & Avg. $\uparrow$ \\
        \midrule
        1Block & 3.73 & 13.39 & 1.77 & 6.30 \\
        2Block & 5.72 & 14.52 & 4.71 & 8.32 \\
        Full DBS (4 blocks) & \textbf{6.48} & \textbf{15.01} & \textbf{5.73} & \textbf{9.07} \\
        \bottomrule
    \end{tabular}
\end{table}

\textbf{Boundary prediction.} To assess the contribution of learned spectral features to boundary prediction, we evaluate a magnitude-only predictor that generates boundary scores from magnitude statistics without learned spectral features (Table~\ref{tab:functional_ablation}). Average SI-SDR decreases from 9.07 to 6.93~dB, a drop of 2.14~dB. This result supports using learned spectral features alongside magnitude statistics for boundary prediction.

\begin{table}[htbp]
    \centering
    \caption{Boundary prediction and frequency reconstruction ablations for DBS-BSRNN on the DnR v2 test set. Magnitude-only predicts boundaries from magnitude statistics without learned spectral features. Rows marked ``w/o'' remove the named component from full DBS.}
    \label{tab:functional_ablation}
    \small
    \setlength{\tabcolsep}{6pt}
    \begin{tabular}{lrrrr}
        \toprule
        Variant & Music $\uparrow$ & Speech $\uparrow$ & SFX $\uparrow$ & Avg. $\uparrow$ \\
        \midrule
        \textbf{Full DBS} & \textbf{6.48} & \textbf{15.01} & \textbf{5.73} & \textbf{9.07} \\
        \midrule
        Magnitude-only predictor & 4.33 & 14.02 & 2.43 & 6.93 \\
        w/o explicit energy & 5.88 & 14.50 & 5.11 & 8.50 \\
        w/o acoustic transition prior & 0.49 & 12.83 & $-1.04$ & 4.09 \\
        w/o confidence path & 5.06 & 13.90 & 3.62 & 7.53 \\
        \midrule
        w/o frequency refinement & 0.11 & 13.26 & $-1.61$ & 3.92 \\
        \bottomrule
    \end{tabular}
\end{table}

We further test how statistical inputs, the boundary-score prior, and the gradient path affect separation. Removing explicit energy descriptors (the predictor's magnitude-statistic inputs) reduces average SI-SDR by 0.57~dB. Removing the acoustic transition prior, which adds adjacent-bin changes in magnitude statistics to the boundary score, reduces it by 4.98~dB. Removing the confidence path for boundary gradients reduces it by 1.54~dB. These results support retaining all three components, particularly the transition prior.

\textbf{Frequency reconstruction.} To assess the role of frequency refinement after token expansion, we remove this stage (Table~\ref{tab:functional_ablation}). Average SI-SDR falls from 9.07 to 3.92~dB, a drop of 5.15~dB, with the largest losses in music and sound effects. This result supports refining the expanded features before full-resolution mask prediction.


\section{Conclusion}

We introduced Dynamic Band Splitting (DBS), which adapts frequency boundaries to each mixture under a fixed band-token budget. DBS-BSRNN improved separation on DnR v2, and DBS variants improved the evaluated band-split backbones on MUSDB18-HQ and EchoSet. On DnR v2, matched boundary substitution and fixed-boundary training controls showed that boundary placement affected separation quality. Block and component ablations further identified the roles of per-block budgets, boundary scoring, and frequency refinement. These results show that allocating a fixed frequency-resolution budget according to the mixture can improve separation without lengthening the separator's token sequence.

\textbf{Limitations.} DBS kept the number of bands fixed and selected one partition per inference window, so it did not adapt representation size or boundary positions within a window. Preliminary learnable-count variants tended to select the largest admissible per-block budget, suggesting that variable-budget routing requires an explicit complexity constraint.



\bibliography{iclr2027_conference,additional_references}
\bibliographystyle{iclr2027_conference}

\appendix
\section{Boundary Predictor}
\label{app:boundary_predictor}

Figure~\ref{fig:dbs_boundary_predictor} in the main text shows the predictor; this section specifies its feature extraction and scoring operations.

For block $r$, we use $\mathbf{X}_r=\mathbf{X}[\mathcal{F}_r,:]\in\mathbb{C}^{F_r\times T}$ as in Section~\ref{sec:dynamic_partition} and stack its real and imaginary parts into $\mathbf{Y}_r=\operatorname{Stack}_{\mathrm{ch}}(\operatorname{Re}(\mathbf{X}_r),\operatorname{Im}(\mathbf{X}_r))\in\mathbb{R}^{F_r\times T\times2}$. The predictor computes
\begin{equation}
    \begin{aligned}
        \mathbf{C}_r&=\mathcal{C}_r(\mathbf{Y}_r)\in\mathbb{R}^{F_r\times T'\times d},\\
        \mathbf{A}_r&=\mathcal{A}(|\mathbf{X}_r|)\in\mathbb{R}^{F_r\times3},\\
        \mathbf{H}^{\mathrm{bnd}}_r&=\mathcal{R}_r(\mathbf{C}_r,\mathbf{A}_r)\in\mathbb{R}^{F_r\times d}.
    \end{aligned}
    \label{eq:boundary_features}
\end{equation}
Here, $T'$ is the downsampled temporal length and $d$ is the convolutional output width. The convolutional branch $\mathcal{C}_r$ extracts spectro-temporal features, $\mathcal{A}$ computes per-bin magnitude statistics, and the BiLSTM $\mathcal{R}_r$ fuses them. The per-bin statistics vector is $\mathbf{a}_r(f)=\mathbf{A}_r[f,:]^{\top}\in\mathbb{R}^{3}$; the contextual vector from the main text is $\mathbf{h}_r(f)=\mathbf{H}^{\mathrm{bnd}}_r[f,:]^{\top}\in\mathbb{R}^{d}$.

\textbf{Convolutional features.}
The convolutional branch $\mathcal{C}_r$ has two stages:
\begin{equation}
    \begin{aligned}
        \mathbf{C}^{(1)}_r
        &=\operatorname{GELU}\!\left(
            \operatorname{GN}_1\!\left(
                \operatorname{Conv}_1(\mathbf{Y}_r)
            \right)
        \right)
        \in\mathbb{R}^{F_r\times T_1\times(d/2)},\\
        \mathbf{C}_r
        &=\operatorname{GELU}\!\left(
            \operatorname{GN}_2\!\left(
                \operatorname{Conv}_2(\mathbf{C}^{(1)}_r)
            \right)
        \right)
        \in\mathbb{R}^{F_r\times T'\times d}.
    \end{aligned}
    \label{eq:boundary_convolution}
\end{equation}
Both convolutions use $3\times3$ kernels and stride $(1,2)$ along the frequency and time axes, respectively. Their output widths are $d/2$ and $d$; $T_1$ and $T'$ denote the respective output time lengths. The operators $\operatorname{GN}_1$ and $\operatorname{GN}_2$ denote group normalization. The two stages preserve the frequency axis and downsample time by a total factor of four, retaining a sequence of $T'$ feature vectors at each frequency bin.

\textbf{Magnitude statistics.}
The branch $\mathcal{A}$ computes statistics independently for each frequency bin. Let $m_r(f,t)=|[\mathbf{X}_r]_{f,t}|\in\mathbb{R}_{\geq0}$ be its magnitude and $\bar{m}_r(f)=T^{-1}\sum_{t=1}^{T}m_r(f,t)$ its temporal mean. The vector $\mathbf{a}_r(f)$ is
\begin{equation}
    \mathbf{a}_r(f)
    =
    \left[
        \log\!\left(1+\bar{m}_r(f)\right),
        \log\!\left(1+\frac{\max_t m_r(f,t)}{\bar{m}_r(f)+\epsilon}\right),
        \log\!\left(1+\frac{\operatorname{std}_t m_r(f,t)}{\bar{m}_r(f)+\epsilon}\right)
    \right]^{\top}
    \in\mathbb{R}^{3},
\end{equation}
where $\epsilon>0$ is a scalar numerical stabilizer and $\operatorname{std}_t$ denotes the temporal standard deviation. Thus, the three components encode the temporal mean and the peak-to-mean and standard-deviation-to-mean ratios after logarithmic compression.

\textbf{Feature fusion and recurrent aggregation.}
Let $\mathcal{P}_{r,\mathrm{stat}}:\mathbb{R}^{3}\rightarrow\mathbb{R}^{d/4}$ be the learned statistics projection, applied to each row of $\mathbf{A}_r$. Repeating the projected rows over the $T'$ time positions and concatenating them with $\mathbf{C}_r$ gives
\begin{equation}
    \mathbf{V}_r
    =
    \operatorname{Concat}_{\mathrm{ch}}\!\left(
        \mathbf{C}_r,
        \operatorname{Repeat}_{T'}\!\left(
            \mathcal{P}_{r,\mathrm{stat}}(\mathbf{A}_r)
        \right)
    \right)
    \in\mathbb{R}^{F_r\times T'\times(5d/4)}.
    \label{eq:boundary_fusion}
\end{equation}
Here, $\operatorname{Repeat}_{T'}$ broadcasts the projected statistics along the time axis, and $\operatorname{Concat}_{\mathrm{ch}}$ concatenates along the feature-channel axis. For each frequency bin, the temporal sequence $\mathbf{V}_r[f,:,:]\in\mathbb{R}^{T'\times(5d/4)}$ is processed independently by a single-layer BiLSTM with hidden width $d/2$ per direction. In batched execution, the batch and frequency axes are merged, giving an LSTM input of shape $(B F_r)\times T'\times(5d/4)$ for batch size $B$.

Let $\overrightarrow{\mathbf{h}}_r(f,t'),\overleftarrow{\mathbf{h}}_r(f,t')\in\mathbb{R}^{d/2}$ denote the forward and backward hidden states at temporal index $t'\in\{1,\ldots,T'\}$. The per-bin feature concatenates the two terminal states:
\begin{equation}
    \mathbf{h}_r(f)
    =\operatorname{Concat}_{\mathrm{ch}}\!\left(
        \overrightarrow{\mathbf{h}}_r(f,T'),
        \overleftarrow{\mathbf{h}}_r(f,1)
    \right)
    \in\mathbb{R}^{d}.
    \label{eq:boundary_temporal_readout}
\end{equation}
The forward state is taken after processing frames $1$ through $T'$, and the backward state after processing frames $T'$ through $1$. Stacking $\mathbf{h}_r(f)$ over the frequency bins in the block gives $\mathbf{H}^{\mathrm{bnd}}_r\in\mathbb{R}^{F_r\times d}$, completing $\mathcal{R}_r$ in \Eqref{eq:boundary_features}. Temporal averaging is confined to the mean statistic in $\mathcal{A}$; the convolutional feature sequence is summarized by the BiLSTM terminal states.

\textbf{Boundary Probability.}
Let $\mathcal{Q}_r,\mathcal{K}_r:\mathbb{R}^{d}\rightarrow\mathbb{R}^{d_q}$ denote the learned projection maps. The normalized vectors used in \Eqref{eq:boundary_score} are
\begin{equation}
    \mathbf{q}_r(f)
    =\operatorname{Norm}_2\!\left(\mathcal{Q}_r(\mathbf{h}_r(f))\right),
    \qquad
    \mathbf{k}_r(f)
    =\operatorname{Norm}_2\!\left(\mathcal{K}_r(\mathbf{h}_r(f))\right),
    \label{eq:boundary_projections}
\end{equation}
$\operatorname{Norm}_2$ denotes $\ell_2$ normalization, and $\mathbf{q}_r(f),\mathbf{k}_r(f)\in\mathbb{R}^{d_q}$.
For an internal candidate boundary $f>f_r^{\min}$, the score and calibrated probability are
\begin{equation}
    \begin{aligned}
        u_r(f)
        &=\frac{1-\mathbf{q}_r(f)^{\top}\mathbf{k}_r(f-1)}{2}
        +0.15\tanh\!\left(
            \frac{\|\mathbf{a}_r(f)-\mathbf{a}_r(f-1)\|_1}{3}
        \right),\\
        \ell_r(f)
        &=\tau_r\bigl(u_r(f)-0.5\bigr)+\beta_r(f),
        \qquad p_r(f)=\sigma\!\left(\ell_r(f)\right).
    \end{aligned}
    \label{eq:boundary_score}
\end{equation}
The inner product supplies the contextual similarity term in $u_r(f)$, and the mean absolute difference of $\mathbf{a}_r(f)$ and $\mathbf{a}_r(f-1)$ supplies the statistics term. The coefficient $0.15$ and offset $0.5$ are fixed, while $\tau_r\geq1$ and $\beta_r(f)\in\mathbb{R}$ are learned.

For the first frequency bin, the score is padded with $u_r(f_r^{\min})=1$, but the implementation then overwrites the resulting logit with $10$. Its Boundary Probability is therefore $p_r(f_r^{\min})=\sigma(10)$. The boundary-density regularizer uses this probability, while partition construction fixes $b_r(f_r^{\min})=1$ and selects the remaining $J_r-1$ boundaries by Hard Top-K.

\section{Band Tokenization and Reconstruction Details}

\subsection{Content-Dependent Band Pooling}
\label{app:tokenization_details}

The bin encoder for each block maps the stacked real and imaginary components $\mathbf{Y}_r\in\mathbb{R}^{F_r\times T\times2}$ to $D$-dimensional bin features:
\begin{equation}
    \mathbf{U}_r
    =\mathcal{E}_r(\mathbf{Y}_r)
    \in\mathbb{R}^{F_r\times T\times D}.
\end{equation}
In the BSRNN instantiation, $\mathcal{E}_r$ applies channel-wise normalization and a $1\!\times\!1$ convolution at every time--frequency location. A two-layer network $g_r:\mathbb{R}^{D}\to\mathbb{R}$ with a $\tanh$ nonlinearity scores each bin after temporal averaging:
\begin{equation}
    e_r(f)
    =g_r\!\left(\frac{1}{T}\sum_{t=1}^{T}\mathbf{U}_r(f,t)\right)
    \in\mathbb{R}.
\end{equation}
Scores are shared over time and normalized within each selected band $\mathcal{S}_{r,j}$:
\begin{equation}
    \alpha_{r,j}(f)
    =\frac{\exp e_r(f)}{\displaystyle\sum_{i\in\mathcal{S}_{r,j}}\exp e_r(i)},
    \qquad \alpha_{r,j}(f)\in[0,1],\quad f\in\mathcal{S}_{r,j}.
\end{equation}
These weights produce the band tokens in Section~\ref{sec:dynamic_tokenization}.

\subsection{Smoothing, Refinement, and Mask Prediction}
\label{app:reconstruction_details}

Probability-Guided Smoothing first averages Boundary Probabilities within each selected band:
\begin{equation}
    \bar p_{r,j}
    =\frac{1}{|\mathcal{S}_{r,j}|}
    \sum_{f\in\mathcal{S}_{r,j}}p_r(f)
    \in[0,1].
    \label{eq:compressed_boundary_probability}
\end{equation}
For each separated token $\mathbf{Z}^{\mathrm{sep}}_{r,j}(t)\in\mathbb{R}^{D_{\mathrm{sep}}}$, its smoothed counterpart $\bar{\mathbf{Z}}_{r,j}(t)\in\mathbb{R}^{D_{\mathrm{sep}}}$ is defined recursively by
\begin{equation}
    \begin{aligned}
        \bar{\mathbf{Z}}_{r,1}(t)
        &=\mathbf{Z}^{\mathrm{sep}}_{r,1}(t),\\
        \bar{\mathbf{Z}}_{r,j}(t)
        &=\bar p_{r,j}\mathbf{Z}^{\mathrm{sep}}_{r,j}(t)
        +\left(1-\bar p_{r,j}\right)\bar{\mathbf{Z}}_{r,j-1}(t),
        \qquad j=2,\ldots,J_r.
    \end{aligned}
    \label{eq:token_smoothing}
\end{equation}
Larger $\bar p_{r,j}$ retains more of the current token; smaller values propagate more from the preceding band. Following the hard expansion in \Eqref{eq:hard_token_to_bin}, the residual frequency-refinement block applies a depthwise convolution along frequency, a pointwise projection, and normalization. Its kernel size and depth are reported in Appendix~\ref{app:implementation}.

The frequency-shared decoder $\mathcal{D}_r:\mathbb{R}^{D_{\mathrm{sep}}}\to\mathbb{R}^{K\times2\times2}$ maps each refined bin feature to mask parameters within block $r$:
\begin{equation}
    \begin{aligned}
        \mathbf{H}_r
        &=\mathcal{R}_r^{\mathrm{freq}}(\mathbf{Z}_r^{e})
        \in\mathbb{R}^{F_r\times T\times D_{\mathrm{sep}}},\\
        \boldsymbol{\Theta}_r(f,t)
        &=\mathcal{D}_r\!\left(\mathbf{H}_r(f,t)\right)
        \in\mathbb{R}^{K\times2\times2}.
    \end{aligned}
\end{equation}
For source $k$ and component $c\in\{\mathrm{Re},\mathrm{Im}\}$, the final axis of $\boldsymbol{\Theta}_r(f,t)$ provides the two scalars of a gated mask component:
\begin{equation}
    \widetilde m_{r,k}^{c}(f,t)
    =[\boldsymbol{\Theta}_r(f,t)]_{k,c,1}
    \sigma\!\left([\boldsymbol{\Theta}_r(f,t)]_{k,c,2}\right)
    \in\mathbb{R},
\end{equation}
which defines the pre-projection mask $\widetilde{\mathbf{M}}_{r,k}\in\mathbb{C}^{F_r\times T}$ through $[\widetilde{\mathbf{M}}_{r,k}]_{f,t}=\widetilde m_{r,k}^{\mathrm{Re}}(f,t)+\mathrm{i}\widetilde m_{r,k}^{\mathrm{Im}}(f,t)$. For $K>1$, let $\mathbf{1}_{F_r\times T}\in\mathbb{R}^{F_r\times T}$ denote the all-ones matrix. The source-axis consistency projection is
\begin{equation}
    \mathbf{M}_{r,k}
    =\widetilde{\mathbf{M}}_{r,k}
    -\frac{\sum_{q=1}^{K}\widetilde{\mathbf{M}}_{r,q}-\mathbf{1}_{F_r\times T}}{K};
\end{equation}
for $K=1$, $\mathbf{M}_{r,1}=\widetilde{\mathbf{M}}_{r,1}$. Concatenating masks from all blocks along frequency gives $\mathbf{M}_k\in\mathbb{C}^{F\times T}$ for each source. As in Section~\ref{sec:full_resolution_reconstruction}, $\hat{\mathbf{S}}_k=\mathbf{X}\odot\mathbf{M}_k\in\mathbb{C}^{F\times T}$ is transformed by the inverse STFT to obtain $\hat{\mathbf{s}}_k$.

\section{Straight-Through Gradient for Discrete Assignments}
\label{app:ste}

For frequency position $f$, let $\mathbf{Z}_r^{(1)}(f),\mathbf{Z}_r^{(0)}(f)\in\mathbb{R}^{T\times D_{\mathrm{sep}}}$ collect the smoothed token features over all $T$ frames under the local hypotheses that a boundary is present or absent. With one-based band indices, the two local assignments are
\begin{equation}
    j_r^{(1)}(f)=s_r(f)+1-b_r(f),
    \qquad
    j_r^{(0)}(f)=s_r(f)-b_r(f).
\end{equation}
Thus, the boundary-present hypothesis uses the higher-index local band and the boundary-absent hypothesis uses the lower-index local band. An invalid neighboring index falls back to the hard-selected band. The soft assignment surrogate is the feature matrix
\begin{equation}
    \mathbf{Z}_r^{\mathrm{soft}}(f)
    =
    p_r(f)\mathbf{Z}_r^{(1)}(f)
    +
    \left[1-p_r(f)\right]\mathbf{Z}_r^{(0)}(f)
    \in\mathbb{R}^{T\times D_{\mathrm{sep}}}.
\end{equation}
Let $\mathbf{Z}_r^{e}(f)\in\mathbb{R}^{T\times D_{\mathrm{sep}}}$ be the feature matrix gathered from the selected hard partition, whose $t$th row is $\mathbf{Z}_r^{e}(f,t)^{\top}$. The straight-through representation is
\begin{equation}
    \mathbf{Z}_r^{\mathrm{STE}}(f)
    =
    \mathbf{Z}_r^{e}(f)
    +
    \mathbf{Z}_r^{\mathrm{soft}}(f)
    -
    \operatorname{sg}\!\left(\mathbf{Z}_r^{\mathrm{soft}}(f)\right)
    \in\mathbb{R}^{T\times D_{\mathrm{sep}}},
\end{equation}
where $\operatorname{sg}(\cdot)$ denotes stop-gradient. Thus, the forward value is exactly the hard assignment, while the backward pass differentiates through the neighboring-assignment surrogate. If the alternative band index is outside $\{1,\ldots,J_r\}$, its representation is replaced by that of the selected band and contributes zero assignment gradient.

The confidence of the selected decision is
\begin{equation}
    c_r(f)
    =
    b_r(f)p_r(f)
    +
    \left[1-b_r(f)\right]\left[1-p_r(f)\right]
    \in[0,1].
\end{equation}
We retain the same hard forward value while adding the confidence path through
\begin{equation}
    \mathbf{Z}_r^{\mathrm{conf}}(f)
    =
    \mathbf{Z}_r^{\mathrm{STE}}(f)
    \left[c_r(f)+\operatorname{sg}\!\left(1-c_r(f)\right)\right]
    \in\mathbb{R}^{T\times D_{\mathrm{sep}}}.
\end{equation}
The bracket evaluates to one during the forward pass and supplies a direct gradient to the boundary probabilities during backpropagation.

For the local derivative with respect to $p_r(f)$, hold the hard assignment and candidate feature matrices fixed. The interpolation gives $\partial\mathbf{Z}_r^{\mathrm{soft}}(f)/\partial p_r(f)=\mathbf{Z}_r^{(1)}(f)-\mathbf{Z}_r^{(0)}(f)$, while $\partial c_r(f)/\partial p_r(f)=2b_r(f)-1$. Applying the product rule to the confidence path yields
\begin{equation}
    \frac{\partial\mathbf{Z}_r^{\mathrm{conf}}(f)}{\partial p_r(f)}
    =\mathbf{Z}_r^{(1)}(f)-\mathbf{Z}_r^{(0)}(f)
    +\bigl(2b_r(f)-1\bigr)\mathbf{Z}_r^{e}(f).
    \label{eq:boundary_surrogate_gradient}
\end{equation}
The difference of the candidate features supplies the interpolation gradient, and the last term supplies the confidence contribution. Multiplication by $p_r(f)[1-p_r(f)]$ gives the corresponding local derivative with respect to the boundary logit $\ell_r(f)$. These are surrogate gradients for training; the forward partition remains the Hard Top-K selection.

The boundary-density regularizer in \Eqref{eq:ratio_loss} acts on the soft probability field used by the surrogate and confidence paths; the realized hard band count is fixed separately by Hard Top-$K$.

\section{Source-Separation Objectives}
\label{app:separation_objectives}

DnR v2 and MUSDB18-HQ have named sources, so their training losses matched each output to its corresponding reference. EchoSet has two interchangeable speaker outputs and used permutation-invariant training (PIT). We computed SI-SDR and SNR after mean-centering the estimated waveform $\hat{\mathbf{s}}_k\in\mathbb{R}^{L}$ and its reference $\mathbf{s}_k\in\mathbb{R}^{L}$. In the following score definitions, these symbols denote their centered versions. With the scalar projection coefficient $a_k=\hat{\mathbf{s}}_k^{\top}\mathbf{s}_k/\|\mathbf{s}_k\|_2^2$, the scores were
\begin{equation}
    \begin{aligned}
        \operatorname{SI\text{-}SDR}(\hat{\mathbf{s}}_k,\mathbf{s}_k)
        &=10\log_{10}\frac{\|a_k\mathbf{s}_k\|_2^2}
        {\|\hat{\mathbf{s}}_k-a_k\mathbf{s}_k\|_2^2},\\
        \operatorname{SNR}(\hat{\mathbf{s}}_k,\mathbf{s}_k)
        &=10\log_{10}\frac{\|\mathbf{s}_k\|_2^2}
        {\|\hat{\mathbf{s}}_k-\mathbf{s}_k\|_2^2}.
    \end{aligned}
\end{equation}
Numerical stabilizers are omitted for clarity.

\textbf{DnR v2.} The fixed and DBS BSRNN models minimized mean negative SI-SDR over the three named sources:
\begin{equation}
    \mathcal{L}_{\mathrm{SI\text{-}SDR}}
    =-\frac{1}{K}\sum_{k=1}^{K}
    \operatorname{SI\text{-}SDR}\!\left(\hat{\mathbf{s}}_k,\mathbf{s}_k\right),
    \qquad K=3.
    \label{eq:dnr_si_sdr_loss}
\end{equation}
The fixed and DBS TIGER models followed the cinematic time-frequency MAE objective~\citep{xu2025tiger}. Let $\mathbf{Q}_k=\operatorname{STFT}(\mathbf{s}_k)\in\mathbb{C}^{F\times T}$ and $\hat{\mathbf{Q}}_k=\operatorname{STFT}(\hat{\mathbf{s}}_k)\in\mathbb{C}^{F\times T}$ be the spectra of the reference and estimate. The waveform and real/imaginary spectral errors were
\begin{equation}
    \begin{aligned}
        \mathcal{L}_{\mathrm{TF\text{-}MAE}}
        =\frac{1}{K}\sum_{k=1}^{K}\Bigg[
        &\frac{\|\hat{\mathbf{s}}_k-\mathbf{s}_k\|_1}{L}\\
        &+\frac{
        \|\operatorname{Re}(\hat{\mathbf{Q}}_k-\mathbf{Q}_k)\|_1
        +\|\operatorname{Im}(\hat{\mathbf{Q}}_k-\mathbf{Q}_k)\|_1}{FT}
        \Bigg].
    \end{aligned}
    \label{eq:time_frequency_mae}
\end{equation}
Here, $\|\cdot\|_1$ sums the absolute values of all entries; division by $L$ and $FT$ forms the respective means.

\textbf{MUSDB18-HQ.} The fixed and DBS variants of BSRNN, BS-RoFormer, and MelRoFormer jointly estimated the four named stems (vocals, drums, bass, and other), matching each prediction to its corresponding reference. All three backbone pairs used the framework's waveform MAE plus multi-resolution complex-STFT MAE. The spectral term summed the mean absolute complex-spectrogram error at window lengths $\{4096,2048,1024,512,256\}$, with a weight of 1.0 in the public four-stem configurations. The BSRNN pair followed the BS-RoFormer training configuration.

\textbf{EchoSet.} Let $\Pi_2$ be the set of two possible speaker assignments. DBS-BSRNN minimized PIT negative SI-SDR,
\begin{equation}
    \mathcal{L}_{\mathrm{PIT\text{-}SI\text{-}SDR}}
    =\min_{\pi\in\Pi_2}
    \left[-\frac{1}{2}\sum_{k=1}^{2}
    \operatorname{SI\text{-}SDR}\!\left(\hat{\mathbf{s}}_k,\mathbf{s}_{\pi(k)}\right)\right],
    \label{eq:pit_si_sdr_loss}
\end{equation}
whereas DBS-TIGER used PIT negative SNR,
\begin{equation}
    \mathcal{L}_{\mathrm{PIT\text{-}SNR}}
    =\min_{\pi\in\Pi_2}
    \left[-\frac{1}{2}\sum_{k=1}^{2}
    \operatorname{SNR}\!\left(\hat{\mathbf{s}}_k,\mathbf{s}_{\pi(k)}\right)\right].
    \label{eq:pit_snr_loss}
\end{equation}
Both losses were averaged over the training batch. The TIGER model followed its released training configuration, which specified negative SNR rather than the negative SI-SDR described in the paper~\citep{xu2025tiger}.\footnote{TIGER's public small-model configuration, commit \texttt{9f18d4a}: \url{https://github.com/JusperLee/TIGER/blob/9f18d4a10a7137e1ce8052cfb62215179f1287b6/configs/tiger-small.yml}. The large-model configuration uses the same losses.}

\section{Dataset Details}
\label{app:datasets}

Table~\ref{tab:dataset_details} summarizes the sampling rates, recording durations, and training/test partitions of the three benchmarks.

\textbf{DnR v2.} Divide and Remaster (DnR v2)~\citep{petermann2022cocktail} provides synthetic monaural soundtracks for separating music, speech, and sound effects.\footnote{Dataset download: \url{https://zenodo.org/records/6949108}.} Music clips were drawn from FMA, speech from LibriSpeech, and sound effects from FSD50K. The construction procedure varied source overlap and relative loudness to approximate soundtrack mixing, with foreground events and background ambience combined into the sound-effects stem. Each mixture and its three reference stems were released as 60-second waveforms at 44.1~kHz. The official v2 release contains 3,406 training, 487 validation, and 973 test mixtures. We used the complete training and test partitions and evaluated all 973 full-length test recordings. Its combination of music, speech, and heterogeneous environmental events provides diverse spectral content for studying input-dependent band allocation.

\textbf{MUSDB18-HQ.} MUSDB18-HQ~\citep{rafii2019musdb18hq} is the uncompressed version of MUSDB18 and contains 150 full-length stereo songs spanning multiple genres.\footnote{Dataset download: \url{https://zenodo.org/records/3338373}.} Each track is provided at 44.1~kHz as a mixture and four isolated stems: vocals, drums, bass, and other, with other collecting the remaining accompaniment. The recordings originate primarily from DSD100 and MedleyDB, together with additional contributed tracks. The uncompressed WAV release retains audio bandwidth up to approximately 22~kHz, beyond the approximately 16~kHz bandwidth of the AAC-compressed version. We used the official partition of 100 training tracks and 50 test tracks, without external training audio. Each model was trained to estimate all four stems jointly, and evaluation used the complete test tracks.

\textbf{EchoSet.} EchoSet~\citep{xu2025tiger} targets two-speaker separation with noise, reverberation, and variable temporal overlap.\footnote{Dataset download: \url{https://huggingface.co/datasets/JusperLee/EchoSet}.} LibriSpeech utterances were rendered with SoundSpaces~2.0 in Matterport3D indoor scenes, accounting for room geometry, object occlusion, and acoustic material properties. The two speakers occupied the same room, with source--microphone distances sampled between 1 and 5~m. Their speech signals were combined at randomly sampled overlap ratios and mixed with background noise from WHAM!. The official splits contain 20,268 training, 4,604 validation, and 2,650 test mixtures, each six seconds long at 16~kHz. We used the released training partition and evaluated all 2,650 complete test mixtures. Both targets are speech, and the 16~kHz sampling rate gives an analyzed frequency range of 0--8~kHz, complementing the broader-bandwidth DnR v2 and MUSDB18-HQ benchmarks.

\begin{table}[htbp]
    \centering
    \caption{Dataset sampling rates, recording durations, and official training/test partition sizes. Duration refers to a complete mixture or music track.}
    \label{tab:dataset_details}
    \small
    \begin{tabular}{lrrrr}
        \toprule
        Dataset & Rate (kHz) & Duration (s) & Train & Test \\
        \midrule
        DnR v2 & 44.1 & 60 & 3,406 & 973 \\
        MUSDB18-HQ & 44.1 & variable & 100 & 50 \\
        EchoSet & 16 & 6 & 20,268 & 2,650 \\
        \bottomrule
    \end{tabular}
\end{table}

\section{Model and Training Configurations}
\label{app:implementation}

\subsection{Construction of 128-Band Fixed Partitions}
\label{app:fixed_128_construction}

For the DnR v2 fixed-band controls, we expanded the original layouts of BSRNN~\citep{luo2023bsrnn} and TIGER~\citep{xu2025tiger} independently to 128 bands over 1025 STFT bins. Let $m\in\{\mathrm{B},\mathrm{T}\}$ denote BSRNN or TIGER, respectively, and let $N_m$ be its canonical band count. Its ordered band widths form $\mathbf{w}^{(m)}=(w_1^{(m)},\ldots,w_{N_m}^{(m)})\in\mathbb{Z}_{>0}^{N_m}$. The original layouts were
\begin{equation}
\begin{aligned}
    \mathbf{w}^{(\mathrm{B})}
    &= (2^{\times 24},4^{\times 12},12^{\times 8},24^{\times 8},
       48^{\times 8},128,129), && N_{\mathrm{B}}=62,\\
    \mathbf{w}^{(\mathrm{T})}
    &= (2^{\times 20},4^{\times 10},11^{\times 8},23^{\times 8},
       46^{\times 8},92^{\times 2},121), && N_{\mathrm{T}}=57,
\end{aligned}
\label{eq:canonical_fixed_widths}
\end{equation}
where $a^{\times n}$ denotes $n$ consecutive bands of width $a$. Both width vectors sum to 1025 bins. For each backbone, define the canonical cumulative boundaries $\gamma_0^{(m)}=0$ and $\gamma_i^{(m)}=\sum_{k=1}^{i}w_k^{(m)}$ for $i=1,\ldots,N_m$. We then locate the $j$th of 127 internal boundaries at the equal-rank coordinate
\begin{equation}
    \xi_j^{(m)}=\frac{N_m j}{128},\qquad
    i_j^{(m)}=\lfloor \xi_j^{(m)}\rfloor,\qquad
    \eta_j^{(m)}=\xi_j^{(m)}-i_j^{(m)},
    \quad j=1,\ldots,127,
\end{equation}
and interpolate on the corresponding cumulative-boundary curve:
\begin{equation}
    \widetilde{B}_j^{(m)}
    =\gamma_{i_j^{(m)}}^{(m)}+\eta_j^{(m)}w_{i_j^{(m)}+1}^{(m)}.
\end{equation}
The integer boundaries were constructed in increasing order by
\begin{equation}
\begin{aligned}
    B_0^{(m)}&=0,\qquad B_{128}^{(m)}=1025,\\
    B_j^{(m)}&=\max\!\left(
        B_{j-1}^{(m)}+1,
        \min\!\left(\operatorname{round}(\widetilde{B}_j^{(m)}),
        1025-(128-j)\right)
    \right),\quad j=1,\ldots,127.
\end{aligned}
\label{eq:fixed_128_projection}
\end{equation}
The lower and upper clamps preserved strict ordering and reserved at least one STFT bin for every remaining band. The resulting partitions comprised the intervals $[B_{j-1}^{(m)},B_j^{(m)})$, $j=1,\ldots,128$, preserving distinct backbone-specific frequency priors under the same band budget.

\subsection{Frequency Layouts and Backbone Integration}
\label{app:backbone_configs}

\textbf{Frequency layouts.} The ``(128)'' suffix in Table~\ref{tab:dnr_main} denotes 128 band tokens per input window, not a separator hidden dimension. For DBS, this budget did not fix the internal boundary locations. All DBS backbones used four frequency blocks. On DnR v2, we used a 2048-point STFT with 1025 frequency bins. The zero-based block endpoints were 0, 64, 128, 576, and 1025, and the per-block band budgets were $(J_1,J_2,J_3,J_4)=(32,32,32,32)$. Therefore, $J=\sum_{r=1}^{4}J_r=128$. MUSDB18-HQ also used 1025 frequency bins and a 128-band budget.

On EchoSet, both DBS backbones used a 640-point STFT with 321 frequency bins. The block endpoints were $(0,\,40,\,80,\,168,\,321)$ in zero-based bin indexing, with per-block band budgets $(40,\,10,\,8,\,9)$ and $J=67$ tokens in total. Hard Top-$K$ selected the internal boundaries under these budgets. Thus, the block supports and token counts were shared within each dataset, while boundary locations and band widths remained mixture-dependent. Table~\ref{tab:model_config} summarizes the STFT resolutions and band-token budgets.

\begin{table}[t]
    \centering
    \caption{Dataset-level STFT resolution and band-token budget shared across DBS backbones.}
    \label{tab:model_config}
    \small
    \begin{tabular}{lrrr}
        \toprule
        Dataset & STFT bins & Blocks & Band tokens \\
        \midrule
        DnR v2 & 1025 & 4 & 128 \\
        MUSDB18-HQ & 1025 & 4 & 128 \\
        EchoSet & 321 & 4 & 67 \\
        \bottomrule
    \end{tabular}
\end{table}

\textbf{Backbone integration.} BSRNN~\citep{luo2023bsrnn}, TIGER~\citep{xu2025tiger}, BS-RoFormer~\citep{lu2024bsroformer}, and MelRoFormer~\citep{wang2023melband} used different fixed band interfaces. In each DBS integration, the mixture-conditioned partition, content-weighted band encoding, and full-resolution reconstruction described in Section~\ref{sec:method} replaced that interface and its band-based mask reconstruction. The corresponding temporal and inter-band separator remained the modeling core (Table~\ref{tab:backbone_integration}). The fixed DnR v2 BSRNN and TIGER controls each used the 128-band expansion of their own canonical layout (Appendix~\ref{app:fixed_128_construction}). The expanded BSRNN grid retained its finer low-frequency and coarser high-frequency allocation and also initialized DBS-MelRoFormer on MUSDB18-HQ, whereas the fixed MelRoFormer control retained its overlapping Mel filterbank. The DBS and fixed MelRoFormer results therefore compared complete band interfaces, rather than isolating boundary movement within an otherwise identical front end.

\begin{table*}[t]
    \centering
    \caption{Integration of DBS with four band-split separators. DBS replaced each model's band interface and mask reconstruction while retaining its separator core.}
    \label{tab:backbone_integration}
    \small
    \setlength{\tabcolsep}{4pt}
    \begin{tabular*}{\textwidth}{@{\extracolsep{\fill}}l>{\raggedright\arraybackslash}p{0.46\textwidth}>{\raggedright\arraybackslash}p{0.29\textwidth}}
        \toprule
        Backbone & Original band interface & Retained separator \\
        \midrule
        BSRNN & Fixed non-overlapping bands; per-band input and mask heads & Temporal and inter-band RNNs \\
        TIGER & Fixed band splitter and restoration head & FFI blocks \\
        BS-RoFormer & Fixed non-overlapping bands; multi-band masks & Hierarchical RoPE Transformers \\
        MelRoFormer & Overlapping Mel bands; Mel-band masks & Hierarchical RoPE Transformers \\
        \bottomrule
    \end{tabular*}
\end{table*}

\FloatBarrier

\subsection{Optimization}
\label{app:optimization_settings}

We trained the DBS models on eight NVIDIA GeForce RTX 5090 GPUs. On DnR v2 and EchoSet, training used three-second audio segments, a batch size of four per GPU (32 in total), and Adam with an initial learning rate of $10^{-3}$ and zero weight decay. On MUSDB18-HQ, the BS-RoFormer pair followed the framework's public four-stem configuration\footnote{\url{https://github.com/ZFTurbo/Music-Source-Separation-Training/blob/main/configs/config_musdb18_bs_roformer_384_8_2_485100_sage.yaml}}; the BSRNN pair used the same training settings. The MelRoFormer pair followed its public four-stem configuration.\footnote{\url{https://github.com/ZFTurbo/Music-Source-Separation-Training/blob/main/configs/config_musdb18_mel_band_roformer_all_stems.yaml}} These MUSDB runs used Adam with an initial learning rate of $10^{-5}$ and a batch size of one per GPU; segment lengths were 11 seconds for BSRNN and BS-RoFormer and eight seconds for MelRoFormer. Within each backbone pair, the fixed and DBS variants used the same training data and schedule.

In each DBS model, the boundary predictor, band encoder, retained separator, and full-resolution decoder were optimized jointly under the objective in Section~\ref{sec:optimization}. Hard Top-$K$ fixed the forward partition, while the straight-through path and boundary-density regularizer supplied training signals to the predictor.

The DBS decoder used one frequency-direction refinement block. On DnR v2, its kernel size was 3, the decoder received no direct full-resolution mixture-spectrum features, and the ratio-loss weight was $\lambda_{\mathrm{ratio}}=0.01$. On EchoSet, the ratio-loss weight rose from zero to 0.02 during schedule warm-up.

\section{Evaluation Metrics and Protocols}
\label{app:evaluation_protocol}

\subsection{Waveform Scores}
\label{app:metric_definitions}

Let $\mathbf{u},\mathbf{v}\in\mathbb{R}^{L}$ denote an estimated waveform and its reference, respectively, after the preprocessing required by the evaluation protocol. All scores below are measured in decibels, with larger values indicating better reconstruction; numerical stabilizers are omitted for clarity.

\textbf{SI-SDR and SNR.} The least-squares projection of $\mathbf{u}$ onto the span of $\mathbf{v}$ is $a(\mathbf{u},\mathbf{v})\mathbf{v}$, where $a(\mathbf{u},\mathbf{v})=\mathbf{u}^{\top}\mathbf{v}/\|\mathbf{v}\|_2^2$ is a scalar. SI-SDR measures the energy ratio between this projected target and the residual~\citep{leroux2019sdr}:
\begin{equation}
    \operatorname{SI\text{-}SDR}(\mathbf{u},\mathbf{v})
    =10\log_{10}
    \frac{\|a(\mathbf{u},\mathbf{v})\mathbf{v}\|_2^2}
    {\|\mathbf{u}-a(\mathbf{u},\mathbf{v})\mathbf{v}\|_2^2}.
    \label{eq:si_sdr_definition}
\end{equation}
SNR instead compares the estimate with the reference at its original amplitude scale:
\begin{equation}
    \operatorname{SNR}(\mathbf{u},\mathbf{v})
    =10\log_{10}\frac{\|\mathbf{v}\|_2^2}{\|\mathbf{u}-\mathbf{v}\|_2^2}.
    \label{eq:snr_definition}
\end{equation}
The SI-SDR and SNR training losses in Appendix~\ref{app:separation_objectives} used mean-centered waveforms. The public EchoSet evaluator computes SI-SDR without mean centering and uses the same preprocessing for separated estimates and mixture baselines.\footnote{TIGER's public evaluation implementation, commit \texttt{9f18d4a}: \url{https://github.com/JusperLee/TIGER/blob/9f18d4a10a7137e1ce8052cfb62215179f1287b6/look2hear/metrics/wrapper.py}.}

\textbf{SDR.} The BSS Eval definition allows a finite impulse response filter to account for permitted linear distortion~\citep{vincent2006performance}. Let $\ell$ denote its length, $\mathbf{V}_{\ell}(\mathbf{v})\in\mathbb{R}^{(L+\ell-1)\times\ell}$ the matrix whose columns are delayed, zero-padded copies of $\mathbf{v}$, and $\mathbf{u}_{\ell}\in\mathbb{R}^{L+\ell-1}$ the zero-padded estimate. The projected target is $\mathbf{v}^{\mathrm{proj}}=\mathbf{V}_{\ell}(\mathbf{v})\mathbf{V}_{\ell}(\mathbf{v})^{\dagger}\mathbf{u}_{\ell}\in\mathbb{R}^{L+\ell-1}$, where $\dagger$ denotes the Moore--Penrose pseudoinverse. Then
\begin{equation}
    \operatorname{SDR}(\mathbf{u},\mathbf{v})
    =10\log_{10}
    \frac{\|\mathbf{v}^{\mathrm{proj}}\|_2^2}
    {\|\mathbf{u}_{\ell}-\mathbf{v}^{\mathrm{proj}}\|_2^2}.
    \label{eq:sdr_definition}
\end{equation}
For EchoSet, the public TIGER evaluator uses \texttt{fast\_bss\_eval}~\citep{scheibler2022sdr} with its default filter length $\ell=512$ and selects the source permutation that maximizes mean SDR.

\subsection{Source Matching, Averaging, and Improvement Scores}
\label{app:metric_aggregation}

For a test set of $N$ recordings, let $\mathbf{x}^{(n)},\mathbf{s}^{(n)}_k,\hat{\mathbf{s}}^{(n)}_k\in\mathbb{R}^{L_n}$ denote the mixture, reference source, and source estimate for recording $n\in\{1,\ldots,N\}$, where $L_n$ is its length in samples. DnR v2 and MUSDB18-HQ have named output stems, so we scored each estimate against its corresponding reference. For a scalar waveform metric $\mathcal{M}$, the source-wise and overall averages are
\begin{equation}
    \begin{aligned}
        \overline{\mathcal{M}}_k
        &=\frac{1}{N}\sum_{n=1}^{N}
        \mathcal{M}\!\left(\hat{\mathbf{s}}^{(n)}_k,\mathbf{s}^{(n)}_k\right),\\
        \overline{\mathcal{M}}
        &=\frac{1}{K}\sum_{k=1}^{K}\overline{\mathcal{M}}_k.
    \end{aligned}
    \label{eq:metric_average}
\end{equation}
We reported source-wise SI-SDR and its unweighted average for the three DnR v2 sources over $N=973$ recordings.

EchoSet has two unlabeled speaker outputs. Let $\Pi_2$ be the set of their permutations. For each recording and metric, we matched estimates to references using
\begin{equation}
    \pi^{\star}_{n,\mathcal{M}}
    \in\operatorname*{arg\,max}_{\pi\in\Pi_2}
    \sum_{k=1}^{2}
    \mathcal{M}\!\left(\hat{\mathbf{s}}^{(n)}_{\pi(k)},\mathbf{s}^{(n)}_k\right),
    \label{eq:evaluation_matching}
\end{equation}
where $\pi(k)$ indexes the estimate assigned to reference source $k$. Improvement was measured relative to using the unprocessed mixture as the estimate for each target~\citep{xu2025tiger}:
\begin{equation}
    \mathcal{M}_{\mathrm{i}}
    =\frac{1}{NK}\sum_{n=1}^{N}\sum_{k=1}^{K}
    \left[
        \mathcal{M}\!\left(\hat{\mathbf{s}}^{(n)}_{\pi^{\star}_{n,\mathcal{M}}(k)},\mathbf{s}^{(n)}_k\right)
        -\mathcal{M}\!\left(\mathbf{x}^{(n)},\mathbf{s}^{(n)}_k\right)
    \right].
    \label{eq:metric_improvement}
\end{equation}
Setting $\mathcal{M}$ to SDR or SI-SDR gives SDRi or SI-SDRi, respectively, with $K=2$ and $N=2{,}650$. Absolute SI-SDR was also reported for our models where available. Source scores were averaged within each recording, followed by averaging over test recordings.

\subsection{Inference Protocol}
\label{app:test_protocol}

We performed inference on complete test recordings. On DnR v2, we processed all 973 recordings in three-second windows with a 1.5-second hop. For each window, DBS predicted a hard frequency partition from the mixture and used it for both band encoding and full-resolution reconstruction.
On MUSDB18-HQ, each model jointly estimated vocals, drums, bass, and other from input chunks. We combined the chunk-level estimates by overlap-add to reconstruct the four full-length stems.
On EchoSet, we processed each complete six-second mixture in a single forward pass to estimate both speakers.

\section{EchoSet Separation Results}
\label{app:echoset_results}

Table~\ref{tab:echoset_main} reports separation quality on EchoSet. Relative to the published BSRNN scores, DBS-BSRNN improved SDRi by 0.46~dB and SI-SDRi by 0.43~dB. DBS-TIGER exceeded the published TIGER scores by 0.33 and 0.31~dB on the same metrics, respectively. Improvements with both backbones showed that the DBS interface also benefited speech separation within EchoSet's 0--8~kHz analyzed range.

\begin{table*}[h]
    \centering
    \caption{SDRi and SI-SDRi (dB), parameter counts, and MACs per second of input audio on EchoSet. Gray rows mark our DBS results; red subscripts show gains (dB) over the paired baseline.}
    \label{tab:echoset_main}
    \small
    \setlength{\tabcolsep}{7pt}
    \resizebox{\textwidth}{!}{%
        \begin{tabular}{lrrrr}
            \toprule
            Method & SDRi $\uparrow$ & SI-SDRi $\uparrow$
            & Params (M) $\downarrow$ & MACs (G/s) $\downarrow$ \\
            \midrule
            A-FRCNN-16~\citep{hu2021afrcnn} & 9.64 & 8.76 & 6.1 & 81.3 \\
            TDANet Large~\citep{li2023tdanet} & 10.14 & 9.21 & 2.3 & 9.2 \\
            TF-GridNet~\citep{wang2023tfgridnet} & 13.73 & 12.85 & 14.4 & 323.8 \\
            \midrule
            BSRNN~\citep{luo2023bsrnn} & 12.75 & 12.23 & 26.0 & 98.7 \\
            \rowcolor{dbsgray}
            DBS-BSRNN & \dbsgain{13.21}{0.46} & \dbsgain{12.66}{0.43} & 7.1 & 38.1 \\
            \midrule
            TIGER~\citep{xu2025tiger}
            & 14.22 & 13.73 & 0.8 & 15.3 \\
            \rowcolor{dbsgray}
            DBS-TIGER & \dbsgain{14.55}{0.33} & \dbsgain{14.04}{0.31} & 1.1 & 21.1 \\
            \bottomrule
        \end{tabular}%
    }
\end{table*}

\section{Inference-Time Boundary Substitution}
\label{app:boundary_substitution}

We replaced the predicted boundaries of the trained DBS-BSRNN model on DnR v2 with uniform or BSRNN-layout boundaries while keeping model weights, input windows, and the 128-band budget fixed. The learned partition improved average SI-SDR by 1.27 and 0.45~dB over these fixed partitions, respectively (Table~\ref{tab:partition_ablation}). Within the trained model, boundary placement alone therefore affected separation quality.

\begin{table}[h]
    \centering
    \caption{DnR v2 test-set SI-SDR (dB) under matched inference-time boundary substitution. Avg. is the unweighted source mean.}
    \label{tab:partition_ablation}
    \footnotesize
    \setlength{\tabcolsep}{4pt}
    \begin{tabular}{lrrrr}
        \toprule
        Boundary source
        & Music $\uparrow$
        & Speech $\uparrow$
        & SFX $\uparrow$
        & Avg. $\uparrow$ \\
        \midrule
        Uniform & 5.07 & 13.31 & 5.02 & 7.80 \\
        BSRNN layout & 6.04 & 14.36 & 5.46 & 8.62 \\
        Learned DBS
        & \textbf{6.48}
        & \textbf{15.01}
        & \textbf{5.73}
        & \textbf{9.07} \\
        \bottomrule
    \end{tabular}
\end{table}

\section{Diagnostic Analysis of the Learnable Filterbank Baseline}
\label{app:learnable_frontend_diagnostics}

\subsection{Identity Reconstruction Test}

We examined the 128-group learnable filterbank used with BSRNN on DnR v2 at 44.1~kHz, with a 2048-sample analysis window and 512-sample hop. To separate front-end reconstruction from source separation, we bypassed the mask estimator and evaluated analysis followed immediately by synthesis on the same six-second mixed probe. The STFT and inverse STFT underlying the Mel-band control reconstructed the probe accurately, whereas the learned \texttt{Conv1d}/\texttt{ConvTranspose1d} pair produced a waveform with low input correlation and strongly amplified RMS (Table~\ref{tab:learnable_identity}). The analysis--synthesis path therefore introduced substantial distortion independently of the separator output.

\begin{table}[h]
    \centering
    \caption{Identity reconstruction on the same mixed-signal probe with the separator bypassed. SNR is measured in dB; $\rho$ is waveform correlation.}
    \label{tab:learnable_identity}
    \small
    \begin{tabular}{lrrr}
        \toprule
        Front end & SNR & $\rho$ & RMS ratio \\
        \midrule
        STFT (Mel-band control) & 133.6 & 1.000 & 1.00 \\
        Learnable filterbank (K128) & -42.5 & 0.030 & 133.65 \\
        \bottomrule
    \end{tabular}
\end{table}

\subsection{Learned Filter Distribution}

\begin{figure}[h]
    \centering
    \includegraphics[width=0.88\linewidth]{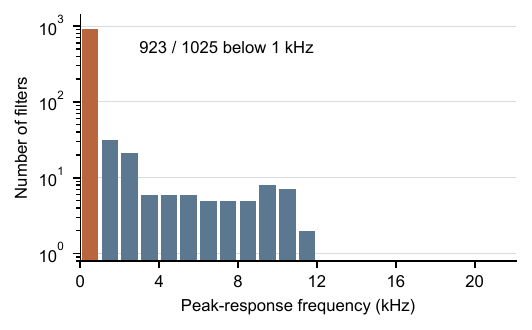}
    \caption{Peak-response frequencies of the 1025 paired analysis filters in the K128 learnable front end. Counts use a logarithmic vertical scale.}
    \label{fig:learnable_filter_peaks}
\end{figure}

The learned analysis filters concentrated their peak responses at low frequencies (Figure~\ref{fig:learnable_filter_peaks}). Of 1025 paired filters, 923 peaked below 1~kHz, and the highest peak was at 11.35~kHz, well below the 22.05~kHz Nyquist frequency. Their magnitude responses were also redundant: mean pairwise cosine similarity was 0.732, with 3,290 pairs above 0.9. This distribution left limited peak-response coverage at higher frequencies.

\subsection{Compatibility with Band-Split Architecture}

BSRNN models neighboring bands as adjacent intervals along the frequency axis. In this implementation, \texttt{StaticBandSplit} grouped each latent-channel pair $(2b,2b+1)$ as though it were frequency bin $b$ and then applied the 128-group layout. The learned waveform filters had no ordering constraint; 80 adjacent filter indices had peak frequencies that reversed by more than 1~kHz. Consequently, the band sequence supplied to intra-band and inter-band modeling did not consistently follow physical frequency order. The model combined a poorly reconstructing analysis--synthesis path with latent ordering inconsistent with the assumed frequency topology.

\section{Boundary Analysis}
\label{app:boundary_analysis}

\subsection{Analysis Protocol and Per-Block Statistics}

We analyzed the boundary behavior of the DnR v2 DBS-BSRNN checkpoint using 64 test mixtures sampled without replacement with a fixed seed. From the central 12 seconds of each mixture, we evaluated seven three-second windows with a 1.5-second hop, giving 448 windows in total. Cross-mixture statistics used one central window per mixture, giving 2,016 distinct pairs. Temporal statistics compared the six adjacent-window pairs within each mixture, giving 384 pairs. We also repeated inference on each central window to test whether identical inputs produced identical boundaries.

For two internal-boundary sets $A$ and $B$ in block $r$, each with $J_r-1=31$ boundaries, we define the boundary replacement rate as
\begin{equation}
    d(A,B) = 1 - \frac{|A \cap B|}{J_r-1}.
\end{equation}
The mandatory start of each block is excluded. A value of zero denotes identical internal boundaries, whereas larger values indicate that more of the fixed boundary budget has moved to different frequency bins.

\begin{figure*}[h]
    \centering
    \includegraphics[width=0.86\textwidth]{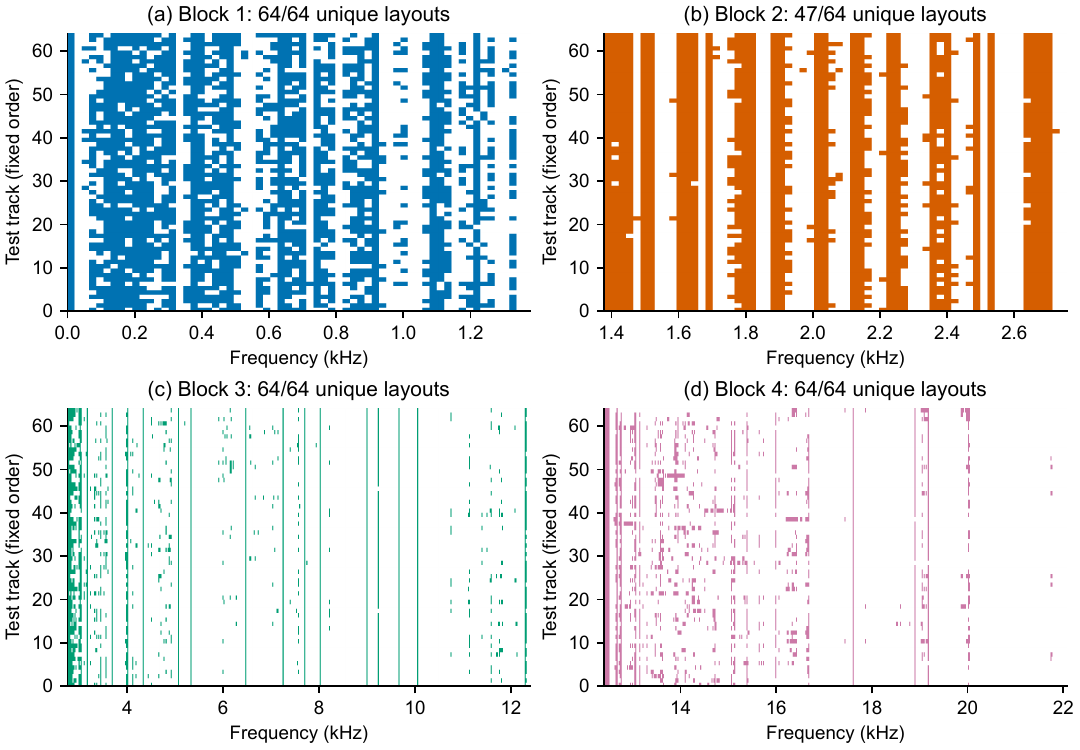}
    \caption{
        Input-dependent frequency partitions learned by DBS-BSRNN on DnR v2.
        Each row corresponds to one test mixture and shows the hard
        boundaries predicted for its central three-second window. Mixture order
        is shared by all panels. Boundary locations varied most strongly in B4,
        whereas B2 retained a comparatively stable layout.
    }
    \label{fig:boundary_raster}
\end{figure*}

\textbf{Variation across mixtures.} Figure~\ref{fig:boundary_raster} shows the hard partitions predicted for the central windows. The complete four-block partition was distinct for all 64 mixtures. B1, B2, B3, and B4 produced 64, 47, 64, and 64 unique partitions, respectively. Their mean cross-mixture replacement rates were 27.56\%, 7.20\%, 26.41\%, and 46.13\% (Table~\ref{tab:boundary_dynamics}). The largest boundary movement occurred in B4 and the smallest in B2.

\begin{table}[t]
    \centering
    \caption{
        Learned boundary dynamics on DnR v2. Unique per-block partitions were
        counted among 64 central windows. Cross-mixture and adjacent-window
        columns report the mean percentage of internal boundaries replaced;
        mandatory block starts are excluded.
    }
    \label{tab:boundary_dynamics}
    \small
    \setlength{\tabcolsep}{3pt}
    \begin{tabular}{lccc}
        \toprule
        Block
        & \shortstack{Unique per-block\\partitions}
        & \shortstack{Cross-mixture\\replacement (\%)}
        & \shortstack{Adjacent-window\\replacement (\%)} \\
        \midrule
        B1 & 64/64 & 27.56 & 22.13 \\
        B2 & 47/64 &  7.20 &  5.69 \\
        B3 & 64/64 & 26.41 & 19.62 \\
        B4 & 64/64 & 46.13 & 32.61 \\
        \bottomrule
    \end{tabular}
\end{table}

\textbf{Temporal adaptation within a mixture.} Adjacent windows exhibited the same pattern across blocks. Their mean replacement rates were 22.13\%, 5.69\%, 19.62\%, and 32.61\% for B1--B4, respectively. The partition responded to local content within a recording rather than assigning one immutable template to each mixture. The router predicted one partition for an entire analysis window, so the observed adaptation was at the window level rather than the frame level.

\textbf{Fixed budget and deterministic routing.} Every analyzed window used the same budget of $J=128$ dynamic bands and therefore the same 128 band tokens. Repeated inference on each of the 64 central windows produced identical hard boundaries and Boundary Probabilities. The variation therefore reflected deterministic input conditioning rather than stochastic inference or a larger band-token budget for selected inputs.

\subsection{Additional Boundary Visualizations}
\label{app:partition_visuals}

\begin{figure*}[h]
    \centering
    \includegraphics[width=0.94\textwidth]{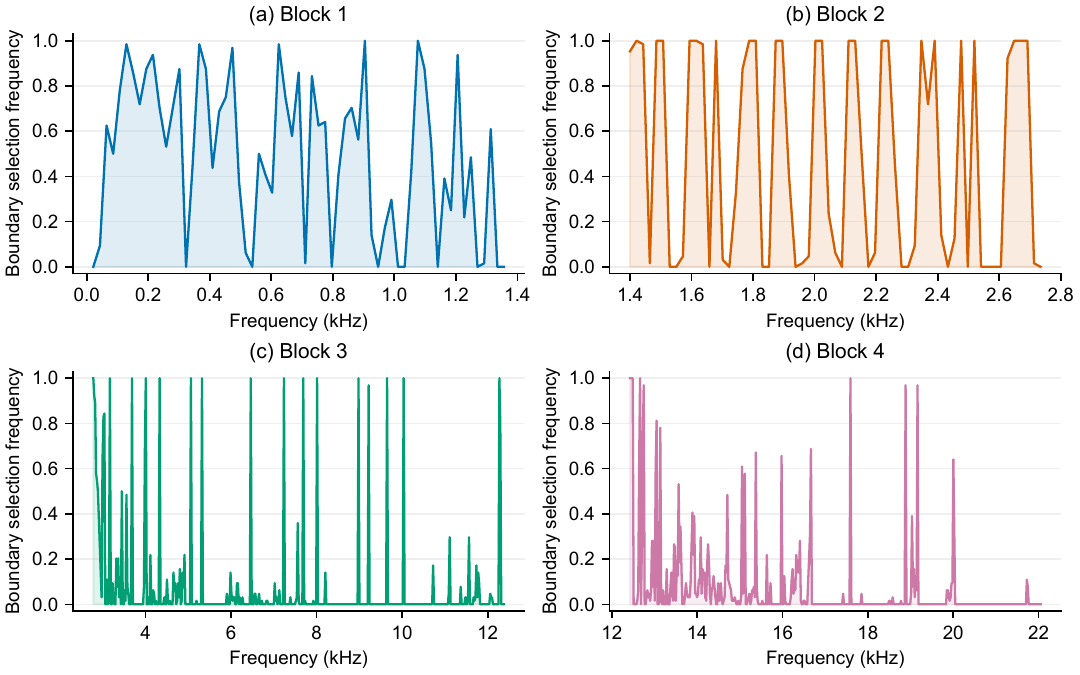}
    \caption{
        Empirical boundary-selection frequency over the central windows of 64
        DnR v2 test mixtures. Mandatory block starts were excluded. B4 had
        the broadest input-dependent selection frequencies, whereas B2 was
        comparatively stable across mixtures.
    }
    \label{fig:boundary_selection_frequency}
\end{figure*}

\textbf{Selection frequency across mixtures.} Figure~\ref{fig:boundary_selection_frequency} summarizes the 64 central windows used for the cross-mixture analysis. For each non-mandatory frequency bin, the plotted value is the fraction of windows in which that bin was selected as an internal boundary. A value of one indicates selection in all analyzed windows rather than a boundary fixed by the block scaffold. A value of zero indicates that the bin was never selected. B2 has many positions with frequencies close to zero or one, consistent with its low cross-mixture replacement rate in Table~\ref{tab:boundary_dynamics}. B1 shows more intermediate selection frequencies, while B3 combines frequently selected positions with many rarely selected candidates. In B4, boundaries are distributed over a wider set of candidate bins, matching its larger cross-mixture replacement rate in Table~\ref{tab:boundary_dynamics}. These patterns show how the same per-block band count can produce different degrees of boundary movement across the frequency axis.

\begin{figure*}[h]
    \centering
    \includegraphics[width=0.90\textwidth]{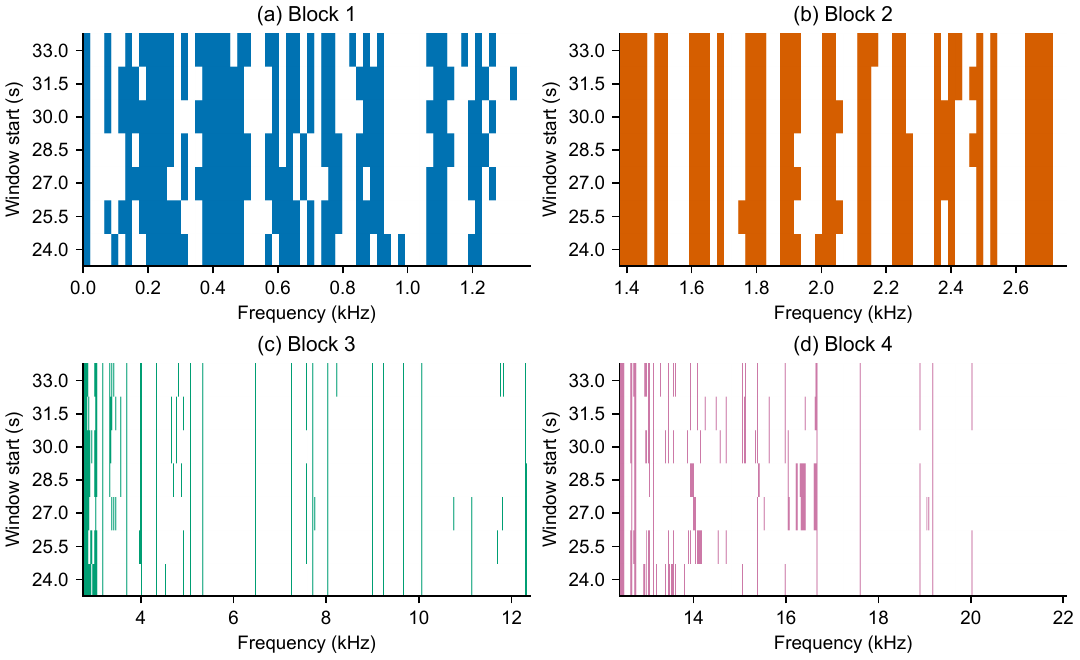}
    \caption{
        Predicted boundaries across seven overlapping windows of one DnR v2 test
        mixture. Each row is a separate 3-second inference window and adjacent
        windows had a 1.5-second hop. The example had the median mean temporal
        replacement rate among the 64 analyzed mixtures. The partition was fixed
        within each window and changed between windows.
    }
    \label{fig:within_track_boundaries}
\end{figure*}

\textbf{Boundary movement within a recording.} Figure~\ref{fig:within_track_boundaries} shows the seven overlapping three-second windows of one test mixture, ordered by window start time. The example was selected as the median track according to its mean adjacent-window replacement rate among the 64 analyzed mixtures. Boundaries that appear at the same horizontal position in successive rows are retained across windows. Offsets and interruptions indicate that a neighboring window selected another frequency bin. Several positions remain aligned through most of the recording, particularly in B2, while other positions change as the analyzed audio advances. B4 shows more such turnover than B2, consistent with their adjacent-window replacement rates of 32.61\% and 5.69\%, respectively, in Table~\ref{tab:boundary_dynamics}. Every row still contains the same per-block boundary budget, so the changing pattern represents a reallocation of boundary locations rather than a change in token count.

\begin{figure*}[h]
    \centering
    \includegraphics[width=0.94\textwidth]{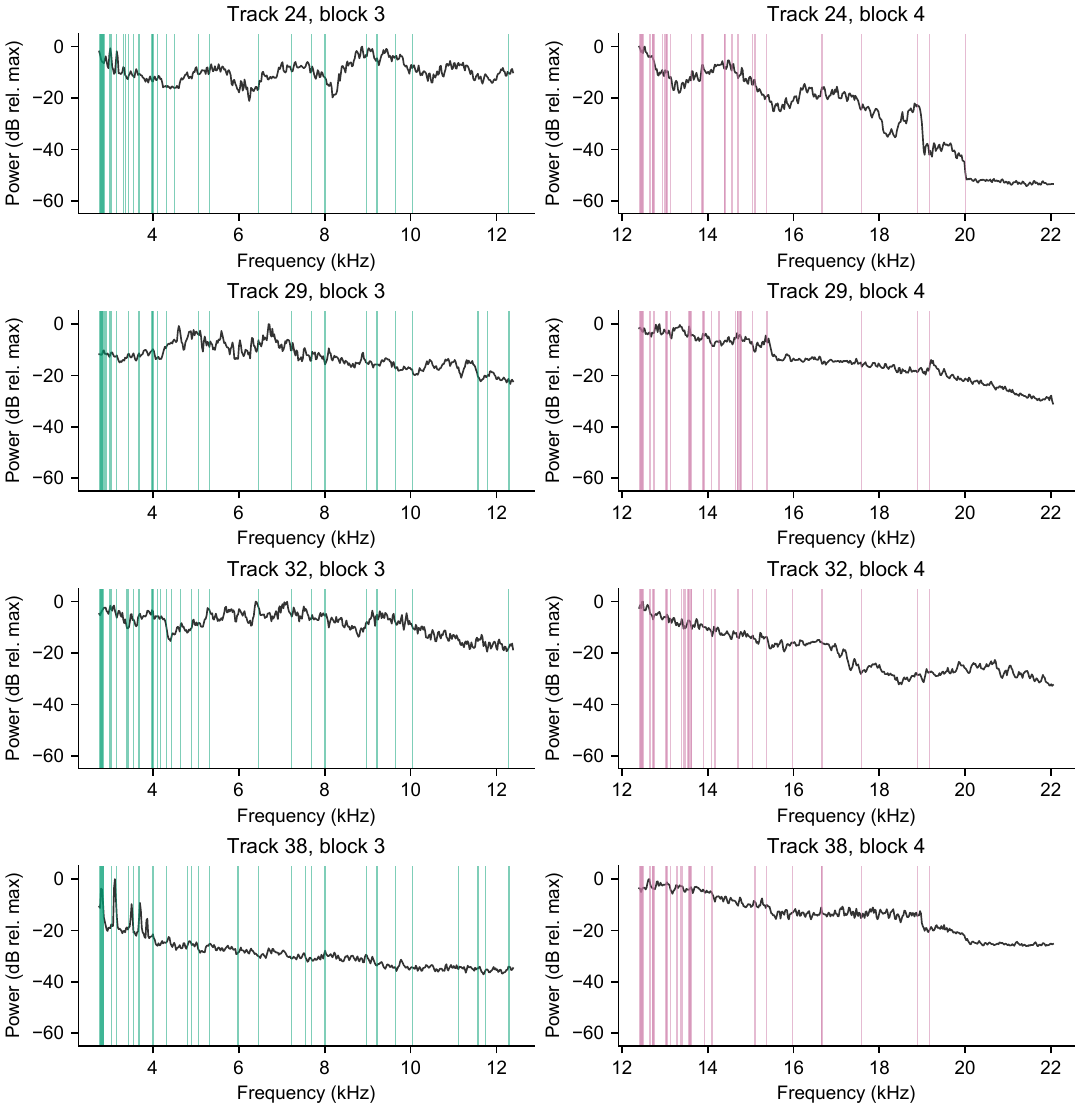}
    \caption{
        Mean input STFT power and predicted hard boundaries for four preselected
        DnR v2 examples in B3 and B4. Power was normalized to the maximum within
        each panel. The visualization supplies input context and does not by
        itself establish semantic alignment or a separation-quality gain.
    }
    \label{fig:spectrum_boundary_examples}
\end{figure*}

\textbf{Boundaries in spectral context.} Figure~\ref{fig:spectrum_boundary_examples} overlays predicted boundaries on the mean mixture STFT power for four preselected examples in B3 and B4. Power is expressed relative to the maximum within each panel, allowing the boundary pattern to be read alongside the shape of that example's spectrum. The lines often form a denser group near the lower edge of a block, but their positions and spacing farther across the block differ between examples. Thus, the fixed budget of 32 bands per block does not imply a common within-block layout. The overlay makes the input-dependent allocation visible alongside mixture-level spectral shape. It does not attribute each selected bin to a particular source or spectral peak.

\end{document}

%% file: math_commands.tex
\usepackage{amsmath,amsfonts,bm}

\def\eqref#1{equation~\ref{#1}}
\def\Eqref#1{Equation~\ref{#1}}

\def\1{\bm{1}}

\DeclareMathAlphabet{\mathsfit}{\encodingdefault}{\sfdefault}{m}{sl}
\SetMathAlphabet{\mathsfit}{bold}{\encodingdefault}{\sfdefault}{bx}{n}



%% file: additional_references.bib
@inproceedings{hu2021afrcnn,
  author    = {Xiaolin Hu and Kai Li and Weiyi Zhang and Yi Luo and Jean-Marie Lemercier and Timo Gerkmann},
  title     = {Speech Separation Using an Asynchronous Fully Recurrent Convolutional Neural Network},
  booktitle = {Advances in Neural Information Processing Systems},
  volume    = {34},
  pages     = {22509--22522},
  year      = {2021},

}

@inproceedings{wang2023tfgridnet,
  author    = {Zhong-Qiu Wang and Samuele Cornell and Shukjae Choi and Younglo Lee and Byeong-Yeol Kim and Shinji Watanabe},
  title     = {{TF-GridNet}: Making Time-Frequency Domain Models Great Again for Monaural Speaker Separation},
  booktitle = {ICASSP 2023 -- 2023 IEEE International Conference on Acoustics, Speech and Signal Processing},
  pages     = {1--5},
  year      = {2023},
}

@article{bengio2013estimating,
  author  = {Yoshua Bengio and Nicholas L{\'e}onard and Aaron Courville},
  title   = {Estimating or Propagating Gradients Through Stochastic Neurons for Conditional Computation},
  journal = {arXiv preprint arXiv:1308.3432},
  year    = {2013},
  
}


%% file: iclr2027_conference.bib
@article{vincent2006performance,
  title   = {Performance Measurement in Blind Audio Source Separation},
  author  = {Vincent, Emmanuel and Gribonval, R{\'e}mi and F{\'e}votte, C{\'e}dric},
  journal = {IEEE Transactions on Audio, Speech, and Language Processing},
  volume  = {14},
  number  = {4},
  pages   = {1462--1469},
  year    = {2006},
}

@inproceedings{scheibler2022sdr,
  title     = {{SDR}---Medium Rare with Fast Computations},
  author    = {Scheibler, Robin},
  booktitle = {2022 IEEE International Conference on Acoustics, Speech and Signal Processing (ICASSP)},
  year      = {2022},
  publisher = {IEEE},
}

@inproceedings{leroux2019sdr,
  title        = {{SDR}---Half-Baked or Well Done?},
  author       = {Le Roux, Jonathan and Wisdom, Scott and Erdogan, Hakan and Hershey, John R.},
  booktitle    = {Proceedings of the IEEE International Conference on Acoustics, Speech and Signal Processing},
  pages        = {626--630},
  year         = {2019},
}

@article{luo2023bsrnn,
  title        = {Music Source Separation With Band-Split {RNN}},
  author       = {Luo, Yi and Yu, Jianwei},
  journal      = {IEEE/ACM Transactions on Audio, Speech, and Language Processing},
  volume       = {31},
  pages        = {1893--1901},
  year         = {2023}
}

@inproceedings{lu2024bsroformer,
  title        = {Music Source Separation with Band-Split {RoPE} Transformer},
  author       = {Lu, Wei-Tsung and Wang, Ju-Chiang and Kong, Qiuqiang and Hung, Yun-Ning},
  booktitle    = {Proceedings of the {IEEE} International Conference on Acoustics, Speech and Signal Processing},
  pages        = {481--485},
  year         = {2024},
  publisher    = {IEEE}
}

@article{wang2023melband,
  title         = {Mel-Band {RoFormer} for Music Source Separation},
  author        = {Wang, Ju-Chiang and Lu, Wei-Tsung and Won, Minz},
  journal       = {arXiv preprint arXiv:2310.01809},
  year          = {2023},
  eprint        = {2310.01809},
  archivePrefix = {arXiv},
  primaryClass  = {cs.SD},
}

@inproceedings{xu2025tiger,
  title        = {{TIGER}: Time-Frequency Interleaved Gain Extraction and Reconstruction for Efficient Speech Separation},
  author       = {Xu, Mohan and Li, Kai and Chen, Guo and Hu, Xiaolin},
  booktitle    = {The Thirteenth International Conference on Learning Representations},
  year         = {2025},
}

@article{luo2019convtasnet,
  title   = {Conv-{TasNet}: Surpassing Ideal Time-Frequency Magnitude Masking for Speech Separation},
  author  = {Luo, Yi and Mesgarani, Nima},
  journal = {IEEE/ACM Transactions on Audio, Speech, and Language Processing},
  year    = {2019}
}

@article{defossez2019demucs,
  title         = {Music Source Separation in the Waveform Domain},
  author        = {D{\'e}fossez, Alexandre and Usunier, Nicolas and Bottou, L{\'e}on and Bach, Francis},
  journal       = {arXiv preprint arXiv:1911.13254},
  year          = {2019},
  eprint        = {1911.13254},
  archivePrefix = {arXiv},
  primaryClass  = {cs.SD},
}

@article{stoter2019openunmix,
  title   = {Open-Unmix---A Reference Implementation for Music Source Separation},
  author  = {St{\"o}ter, Fabian-Robert and Uhlich, Stefan and Liutkus, Antoine and Mitsufuji, Yuki},
  journal = {Journal of Open Source Software},
  year    = {2019}
}

@article{rafii2018overview,
  title   = {An Overview of Lead and Accompaniment Separation in Music},
  author  = {Rafii, Zafar and Liutkus, Antoine and St{\"o}ter, Fabian-Robert and Mimilakis, Stylianos I. and Fitzgerald, Derry and Pardo, Bryan},
  journal = {IEEE/ACM Transactions on Audio, Speech, and Language Processing},
  volume  = {26},
  number  = {8},
  pages   = {1307--1335},
  year    = {2018},
}

@inproceedings{raj2021meeting,
  title     = {Integration of Speech Separation, Diarization, and Recognition for Multi-Speaker Meetings: System Description, Comparison, and Analysis},
  author    = {Raj, Desh and Denisov, Pavel and Chen, Zhuo and Erdogan, Hakan and Huang, Zili and He, Maokui and Watanabe, Shinji and Du, Jun and Yoshioka, T. and Luo, Yi and Kanda, Naoyuki and Li, Jinyu and Wisdom, Scott and Hershey, J.},
  booktitle = {2021 {IEEE} Spoken Language Technology Workshop},
  year      = {2021}
}

@article{li2025advances,
  title={Advances in speech separation: Techniques, challenges, and future trends},
  author={Li, Kai and Chen, Guo and Sang, Wendi and Luo, Yi and Chen, Zhuo and Wang, Shuai and He, Shulin and Wang, Zhong-Qiu and Li, Andong and Wu, Zhiyong and others},
  journal={arXiv preprint arXiv:2508.10830},
  year={2025}
}

@inproceedings{li2023tdanet,
  title         = {An Efficient Encoder-Decoder Architecture with Top-Down Attention for Speech Separation},
  author        = {Li, Kai and Yang, Runxuan and Hu, Xiaolin},
  booktitle     = {The Eleventh International Conference on Learning Representations},
  year          = {2023},
  eprint        = {2209.15200},
  archivePrefix = {arXiv},
  primaryClass  = {cs.SD},
}

@article{wang2023tf,
  title={TF-GridNet: Integrating full-and sub-band modeling for speech separation},
  author={Wang, Zhong-Qiu and Cornell, Samuele and Choi, Shukjae and Lee, Younglo and Kim, Byeong-Yeol and Watanabe, Shinji},
  journal={IEEE/ACM Transactions on Audio, Speech, and Language Processing},
  volume={31},
  pages={3221--3236},
  year={2023},
  publisher={IEEE}
}

@inproceedings{li2026semantically,
  title     = {A Semantically Consistent Dataset for Data-Efficient Query-Based Universal Sound Separation},
  author    = {Li, Kai and Cheng, Jintao and Zeng, Chang and Yan, Zijun and Wang, Helin and Su, Zixiong and Zheng, Bo and Hu, Xiaolin},
  booktitle = {Proceedings of the International Conference on Machine Learning},
  year      = {2026},
}

@inproceedings{petermann2022cocktail,
  title     = {The Cocktail Fork Problem: Three-Stem Audio Separation for Real-World Soundtracks},
  author    = {Petermann, Darius and Wichern, Gordon and Wang, Zhong-Qiu and Le Roux, Jonathan},
  booktitle = {2022 IEEE International Conference on Acoustics, Speech and Signal Processing (ICASSP)},
  pages     = {526--530},
  year      = {2022},
  publisher = {IEEE}
}

@misc{rafii2019musdb18hq,
  title  = {{MUSDB18-HQ} - an uncompressed version of {MUSDB18}},
  author = {Rafii, Zafar and Liutkus, Antoine and St{\"o}ter, Fabian-Robert and Mimilakis, Stylianos Ioannis and Bittner, Rachel},
  year   = {2019},
}

@inproceedings{tong2024scnet,
  title         = {{SCNet}: Sparse Compression Network for Music Source Separation},
  author        = {Tong, Weinan and Zhu, Jiaxu and Chen, Jun and Kang, Shiyin and Jiang, Tao and Li, Yang and Wu, Zhiyong and Meng, Helen M.},
  booktitle     = {2024 IEEE International Conference on Acoustics, Speech and Signal Processing (ICASSP)},
  year          = {2024},
  eprint        = {2401.13276},
  archivePrefix = {arXiv}
}

@article{dai2023learnablefilterbank,
  title         = {Multi-Channel Masking with Learnable Filterbank for Sound Source Separation},
  author        = {Dai, Wang and Politis, A. and Virtanen, Tuomas},
  journal       = {arXiv preprint arXiv:2303.07816},
  year          = {2023},
  eprint        = {2303.07816},
  archivePrefix = {arXiv}
}

@article{saijo2026sfc,
  title         = {Input-Adaptive Spectral Feature Compression by Sequence Modeling for Source Separation},
  author        = {Saijo, Kohei and Bando, Yoshiaki},
  journal       = {IEEE Transactions on Audio, Speech, and Language Processing},
  year          = {2026},
  eprint        = {2602.08671},
  archivePrefix = {arXiv}
}

@article{wang2022sparsemoeseparation,
  title         = {Handling Trade-Offs in Speech Separation with Sparsely-Gated Mixture of Experts},
  author        = {Wang, Xiaofei and Chen, Zhuo and Shi, Yu and Wu, Jian and Kanda, Naoyuki and Yoshioka, Takuya},
  journal       = {arXiv preprint arXiv:2211.06493},
  year          = {2022},
  eprint        = {2211.06493},
  archivePrefix = {arXiv}
}

@article{hu2026tfmoe,
  title         = {{TF-MoE}: Time-Frequency Mixture-of-Experts for Efficient Speech Separation},
  author        = {Hu, Qinzhe and Li, Chenda and Zhang, Wangyou and Liu, Shujie and Lu, Yan and Qian, Yanmin},
  journal       = {arXiv preprint arXiv:2606.29575},
  year          = {2026},
  eprint        = {2606.29575},
  archivePrefix = {arXiv}
}
